\documentclass[amsmath,amssymb,showpacs,twocolumn,superscriptaddress,pr]{revtex4-1}
\usepackage{graphicx,bm,color,subfigure}
\usepackage[T1]{fontenc}
\usepackage{mathrsfs}
\usepackage{bm}
\usepackage{amsmath}
\usepackage{amssymb}
\usepackage{graphicx}
\usepackage{esint}
\usepackage{multirow}
\usepackage{float}
\usepackage{array}
\usepackage{makecell}
\usepackage{harpoon}
\usepackage{booktabs}
\usepackage{gensymb}
\usepackage{simplewick}
\usepackage{subfigure}


\usepackage[english]{babel}

\begin{document}
\title{Kohn-Luttinger mechanism driven exotic topological superconductivities on the Penrose lattice}
\author{Ye Cao}
\affiliation{School of Physics, Beijing Institute of Technology, Beijing 100081, China}
\author{Yongyou Zhang}
\affiliation{School of Physics, Beijing Institute of Technology, Beijing 100081, China}
\author{Yu-Bo Liu}
\affiliation{School of Physics, Beijing Institute of Technology, Beijing 100081, China}
\author{Cheng-Cheng Liu}
\affiliation{School of Physics, Beijing Institute of Technology, Beijing 100081, China}
\author{Wei-Qiang Chen}
\affiliation{Shenzhen Institute for Quantum Science and Engineering and Department of Physics, Southern University of Science and Technology, Shenzhen 518055, China}
\author{Fan Yang}
\email{yangfan_blg@bit.edu.cn}
\affiliation{School of Physics, Beijing Institute of Technology, Beijing 100081, China}
\date{\today}
\begin{abstract}
The Kohn-Luttinger mechanism for unconventional superconductivity (SC) driven by weak repulsive electron-electron interactions on a periodic lattice is generalized to the quasicrystal (QC) via a real-space perturbative approach. The repulsive Hubbard model on the Penrose lattice is studied as an example, on which a classification of the pairing symmetries is performed and a pairing phase diagram is obtained. Two remarkable properties of these pairing states are revealed, due to the combination of the presence of the point-group symmetry and the lack of translation symmetry on this lattice. Firstly, the spin and spacial angular momenta of a Cooper pair is de-correlated: for each pairing symmetry, both spin-singlet and spin-triplet pairings are possible even in the weak-pairing limit. Secondly, the pairing states belonging to the 2D irreducible representations of the $D_5$ point group can be time-reversal-symmetry-breaking topological SCs carrying spontaneous bulk super current and spontaneous vortices. These two remarkable properties are general for the SCs on all QCs, and are rare on periodic lattices. Our work starts the new area of unconventional SCs driven by repulsive interactions on the QC.
\end{abstract}
\pacs{......}

\maketitle
{\bf Introduction:} The quasicrystal (QC) has attracted a lot of research interests \cite{Goldman} since synthesized \cite{Shechtman}. The QC represents a certain type of solid structures which are lack of translation symmetry but can possess rotation symmetries such as the five-folded or eight-folded ones forbidden by crystalline point group \cite{Shechtman}. The electronic structure on a QC is exotic and fundamentally different from that on a crystal. Specifically, due to the lack of translation symmetry on a QC, the lattice momentum is no longer a good quantum number and no Fermi surface (FS) can be defined. Various exotic quantum states with intriguing properties have been revealed on the QC recently \cite{Tsunetsugu1,Tsunetsugu2,Susumu,Wessel,Thiem,Koga,Otsuki,Watanabe,Shaginyan,Takemori,Takemura,Andrade,Kraus,Huang1,Huang2,Longhi,Autti,Giergiel,Lang,Sanchez,Singh,Bandres2016,Hou,Varjas2019,Spurrier2020}. Particularly, the definite experimental evidences for superconductivity (SC) in the recently synthesized Al-Zn-Mg QC \cite{exp}, together with those in previous ternary
QCs \cite{exp2,exp3} and crystalline approximants \cite{exp4}, have attracted a lot of research interests \cite{Sakai2017,theory1,theory2,theory3,attractive}. It's interesting to ask a question here: are there any common features of superconducting states on the QC which are different from those on a crystal?

In Ref \cite{Sakai2017,theory1,attractive}, the pairing states for attractive Hubbard models are studied on QC lattices. It's found that the attractive interactions can lead to Cooper pairing \cite{Cooper_instability} between a time-reversal (TR) partners, obeying the Anderson's theorem \cite{Anderson}. Further more, despite the lack of lattice momentum on the QC, the Cooper pairing can lead to a finite superfluid density \cite{attractive}. These results \cite{Sakai2017,theory1,attractive} suggest that the SC on the QC with attractive interactions is consistent with the BCS theory. However, the situation is distinct for the cases with repulsive interactions, as will be shown below. The pairing in the presence of weak repulsive interactions is induced by the Kohn-Luttinger (KL) mechanism \cite{KL1,KL2}. This theory states that the interaction renormalization brought about by exchanging particle-hole excitations is anisotropic on the FS, which can generate some attractive-interaction channels between the TR partners, which finally leads to Cooper pairing on the FS. Here, we generalize this mechanism to the QC, and obtain unconventional SCs with a series of remarkable properties intrinsic to the QCs which are rare on periodic lattices.

In this paper, we study the KL SC in a weak-U repulsive Hubbard model on a Penrose lattice. Via a real-space perturbative treatment up to the second order, we acquire an effective interaction vertex, through which we derive a linearized gap equation near the superconducting critical temperature $T_c$, solving which we obtain the $T_c$ and the pairing gap functions. We classify the pairing symmetries and obtain the pairing phase-diagram after large scale numerical calculations. Two remarkable results are obtained. Firstly, the orbital- and spin- angular momenta of the Cooper pair are de-correlated even without the spin-orbit coupling (SOC), which means that we can obtain both spin-singlet and spin-triplet pairings for the same pairing symmetry, distinguished from the case on a periodic lattice. Secondly, any 2D irreducible representation (IR) of the $D_5$ point group can bring about TR symmetry (TRS)-breaking topological SCs (TSCs) hosting spontaneous bulk super current and spontaneous vortices. These two properties are caused by the combination of the point-group symmetry and the lack of translation symmetry, and are thus general for the SCs on any QC, and are rare on periodic lattices.
\begin{figure}[htbp]
\centering
\includegraphics[width=0.48\textwidth]{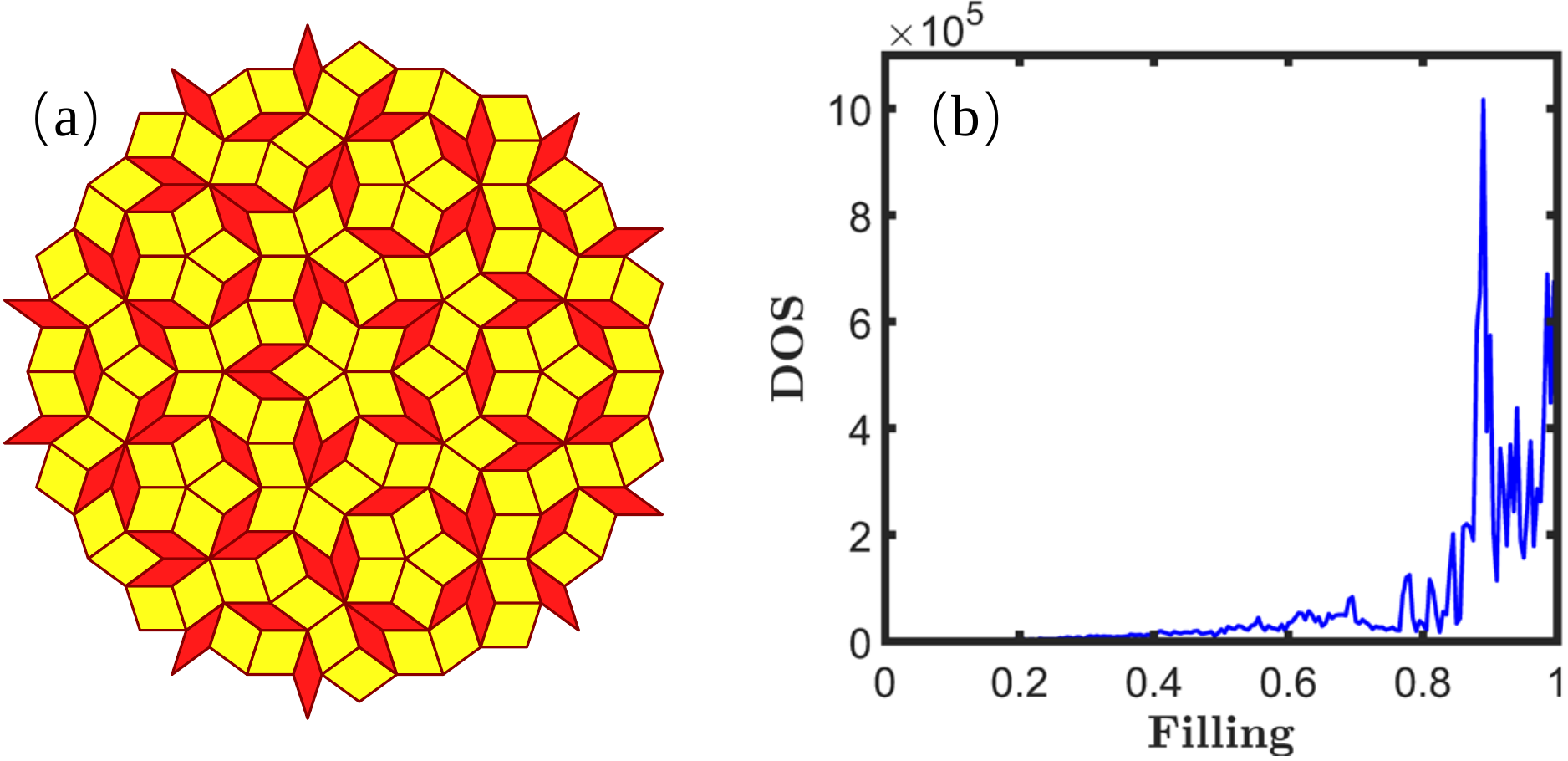}
\caption{\label{fig:demonstration_DOS}(Color online) Lattice pattern with 191 sites (a) and the DOS of the TB Hamiltonian on a lattice with 13926 sites (b). In (a), the lattice constant is $a$.}
\end{figure}

{\bf Model and Approach:} Let's consider the following standard repulsive Hubbard model on the Penrose lattice \cite{Penrose1974} with lattice constant $a$ shown in Fig. 1(a) \cite{footnote1},
\begin{eqnarray}
\hat{\mathcal{H}} &=& -\sum_{\mathbf{i,j},\sigma}t_{\mathbf{ij}}c_{\mathbf{i}\sigma}^{\dagger} c_{\mathbf{j}\sigma} + U\sum_{\mathbf{i}}n_{\mathbf{i}\uparrow}n_{\mathbf{i}\downarrow} -\mu\sum_{\mathbf{i},\sigma}n_{\mathbf{i}\sigma},
\label{rpa_hamiltonian}
\end{eqnarray}
where $c_{\mathbf{i}\sigma}$ annihilates an electron at site $\mathbf{i}$ with spin $\sigma$, $n_{\mathbf{i}\sigma}$ is the electron-number operator, and $\mu$ denotes the chemical potential. The hopping integral $t_{\mathbf{ij}}= e^{-|\mathbf{r_i}-\mathbf{r_j}|/\min(\left\{|\mathbf{r_i}-\mathbf{r_j}|\right\})}$, where $|\mathbf{r_i}-\mathbf{r_j}|$ denotes the distance between different sites $\mathbf{i}$ and $\mathbf{j}$, and $\min(\left\{|\mathbf{r_i}-\mathbf{r_j}|\right\})=0.618a$. The tight-binding (TB) part of Eq.~\eqref{rpa_hamiltonian} is diagonalized as $\hat{\mathcal{H}}_{\text{TB}} = \sum_{m}\tilde{\epsilon}_m c^{\dagger}_{m \sigma}c_{m\sigma}$, with $c_{m \sigma}$ $=$ $\sum_{\mathbf{i}}\xi_{\mathbf{i}m}c_{\mathbf{i}\sigma}$. Here $m$ labels a single-particle eigen state with energy $\tilde{\epsilon}_m=\epsilon_m-\mu$ relative to the chemical potential, and $\xi_{\mathbf{i}m}$ represents for the wave function for the state $m$. The density of states (DOS) at the Fermi energy shown in Fig. 1(b)\cite{footnote2} peaks at around the filling fraction of 0.9, which will be focused on below. In unit of the largest hopping integral, the total band width $W_{D}$ is about 7.56. We consider weak $U>0$ and adopt perturbative approach in our work.

For this repulsive Hubbard model, SC is forbidden in the mean-field (MF) level. However, it can be driven by the KL mechanism, wherein unconventional SC is mediated by exchanging particle-hole excitations. Due to the lack of translation symmetry, we engage a real-space perturbative treatment, whose details are provided in the Supplementary Material (SM) \cite{Supplementary}. The real-space propagator of the particle-hole excitations is described by the susceptibility function, which in the bare level reads \cite{Supplementary}
\begin{eqnarray}\label{chi_0}
\chi_{\mathbf{ij}}^{(0)}(i\Omega_n) &=& \int_{0}^{\beta}e^{i\Omega_{n}\tau} d\tau \langle T_{\tau}c_{\mathbf{i}\uparrow}^{\dagger}(\tau)c_{\mathbf{i}\uparrow}(\tau)c_{\mathbf{j}\uparrow}^{\dagger}c_{\mathbf{j}\uparrow}\rangle_{0}\nonumber\\&=&\sum_{ml}\xi_{\mathbf{i}m}\xi_{\mathbf{j}m}\xi_{\mathbf{i}l}\xi_{\mathbf{j}l} \frac{n_F(\tilde{\epsilon}_m) - n_F(\tilde{\epsilon}_l)}{i\Omega_n + \tilde{\epsilon}_l - \tilde{\epsilon}_m}.
\end{eqnarray}

In our calculations, only about a $1000\times1000$ number of $(m,l)$ near the Fermi level are summed in Eq.~\eqref{chi_0}. In our perturbative treatment \cite{Supplementary}, the four second-order processes of exchanging particle-hole excitations induce effective interactions, from which we obtain \cite{Supplementary}
\begin{eqnarray}\label{heff}
\hat{\mathcal{H}}_{eff} &=& -\sum_{\mathbf{i,j},\sigma}t_{\mathbf{ij}}c_{\mathbf{i}\sigma}^{\dagger} c_{\mathbf{j}\sigma} + U\sum_{\mathbf{i}}n_{\mathbf{i}\uparrow}n_{\mathbf{i}\downarrow} -\mu\sum_{\mathbf{i},\sigma}n_{\mathbf{i}\sigma}\nonumber\\
                  & & -(U^{2}/2)\sum_{\mathbf{i},\mathbf{j},\sigma,\sigma'}\chi_{\mathbf{ij}}c_{\mathbf{i}\sigma}^{\dagger}c_{\mathbf{i}\sigma'}c_{\mathbf{j}\sigma'}^{\dagger}c_{\mathbf{j}\sigma},
\label{eff_hamiltonian}
\end{eqnarray}
with $\chi_{\mathbf{ij}} \equiv \chi^{(0)}_{\mathbf{ij}}(i\Omega_n=0)$. The induced term with coefficient $-(U^{2}/2)$ in Eq.~\eqref{heff} can drive SC in the MF level.

%

\begin{figure}[htbp]
\includegraphics[width=0.48\textwidth]{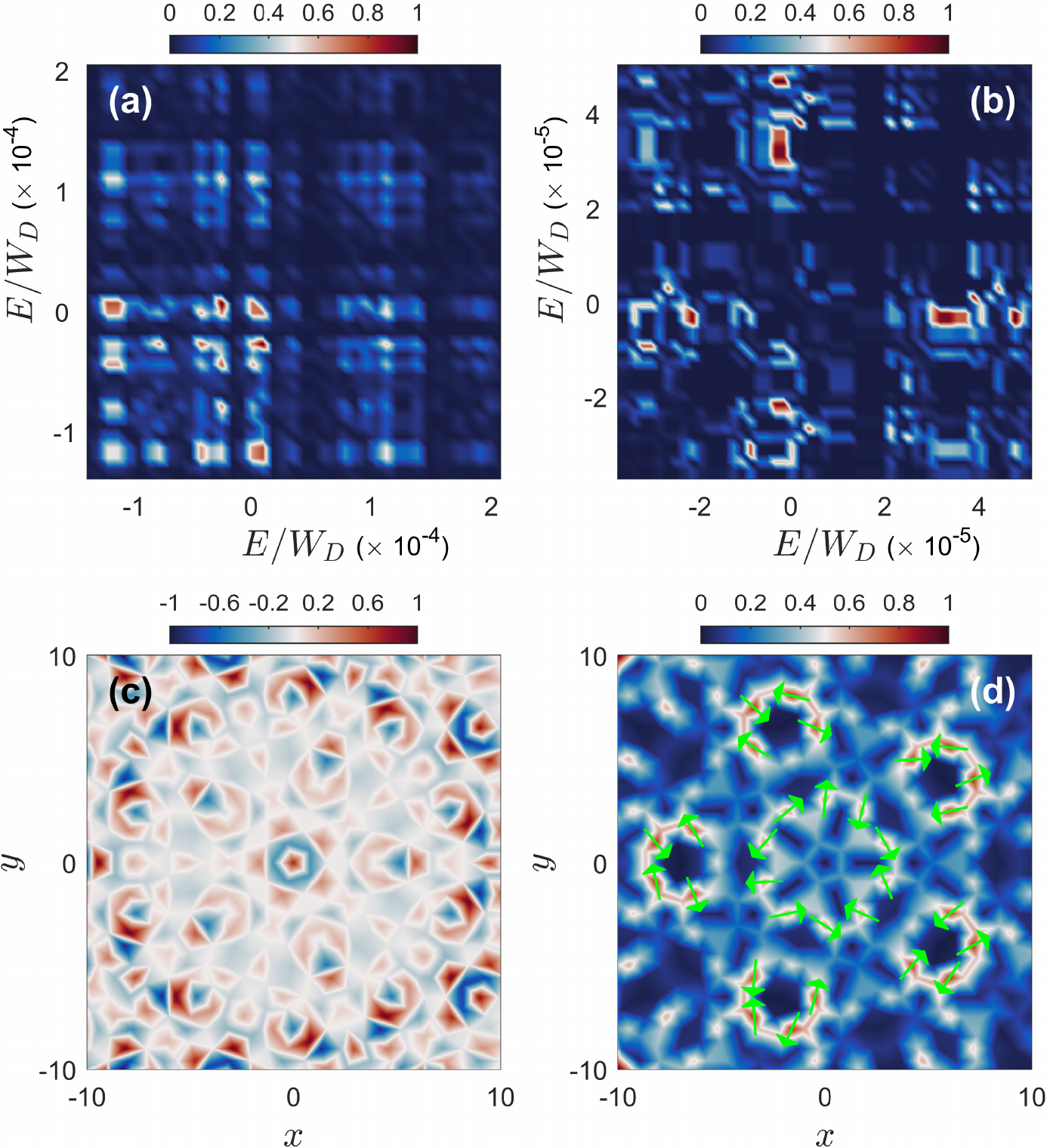}
\caption{\label{fig:Anderson_theorems}(Color online) Contour plots of relative $|\tilde{\Delta}_{mn}|$ and $\Delta_{\mathbf{i},\mathbf{O}}$, where $\mathbf{O}$ is the center of Penrose tiling,  for a singlet $s$ state (a and c, filling=0.81, $U/W_{D}=0.32$) and a triplet $d+id$ state (b  and d,  filling=0.98, $U/W_{D}=0.32$).  In (d), the direction of each marked green arrow at the site $\mathbf{i}$ represents phase angle of $\Delta_{\mathbf{i},\mathbf{O}}$ and the color represents the relative amplitude.}
\end{figure}

\begin{table*}[htbp]
\caption{\label{tab:classification}IRs of the $D_5$ point group and classification of pairing symmetries.  $\hat{R}_{\theta}$ denotes the rotation about the center of the Penrose lattice by the angle $\theta=2n\pi/5$ and $\hat{\sigma}$ represents the reflection about any of the five symmetric axes. $D^{(C_{2\pi/5})}$ and $D^{(\sigma_{x})}$ are the representation matrices for the two generators of $D_5$, i.e. $C_{2\pi/5}$ and $\sigma_{x}$, up to any unitary transformation. For each pairing symmetry listed, both spin-singlet and spin-triplet pairings are possible.}
\begin{tabular}{|p{0.03\textwidth}<{\centering}p{0.03\textwidth}<{\centering}|p{0.18\textwidth}<{\centering}|p{0.07\textwidth}<{\centering}|p{0.20\textwidth}<{\centering}|p{0.32\textwidth}<{\centering}|}
\hline
                         \multicolumn{2}{|c|}{IRs}         &     $D^{(C_{2\pi/5})}$                                      &   $D^{(\sigma_{x})}$                   &  pairing symmetries            & ground-state gap functions\\
\hline
\multirow{2}{*}{1D}      & \multicolumn{1}{|c|}{$A_1$} & $I$        &  $I$                      &  $s$
                         & $\Delta_{\hat{R}_{\theta}\mathbf{i}, \hat{R}_{\theta}\mathbf{j}}=\Delta_{\mathbf{i,j}}$, $\Delta_{\hat{\sigma} \mathbf{i}; \hat{\sigma} \mathbf{j}}=\Delta_{\mathbf{i},\mathbf{j}}$ \\
\cline{2-6}
                         & \multicolumn{1}{|c|}{$A_2$} & $I$         &  $-I$                  &  $h_{y^5-10x^2y^3+5x^4y}$
                         & $\Delta_{\hat{R}_{\theta} \mathbf{i}, \hat{R}_{\theta}\mathbf{j}}=\Delta_{\mathbf{i},\mathbf{j}}$, $\Delta_{\hat{\sigma} \mathbf{i}; \hat{\sigma} \mathbf{j}}=-\Delta_{\mathbf{i},\mathbf{j}}$ \\
\hline
\multirow{2}{*}{2D}      & \multicolumn{1}{|c|}{$E_1$} & $\cos\frac{2\pi}{5} I \pm i\sin\frac{2\pi}{5}\sigma_{y}$                      &  $\sigma_{z}$ & $(p_{x}, p_{y})$
                         & $\Delta_{\hat{R}_{\theta}\mathbf{i}, \hat{R}_{\theta}\mathbf{j}}=e^{\pm i\theta}\Delta_{\mathbf{i},\mathbf{j}}$, $\Delta_{\hat{\sigma} \mathbf{i}, \hat{\sigma} \mathbf{j}} \neq\pm\Delta_{\mathbf{i},\mathbf{j}}$ \\
\cline{2-6}
                         & \multicolumn{1}{|c|}{$E_2$} &  $\cos\frac{4\pi}{5} I \pm i\sin\frac{4\pi}{5}\sigma_{y}$    &  $\sigma_{z}$               &  $(d_{x^2-y^2},d_{2xy})$
                         &$\Delta_{\hat{R}_{\theta}\mathbf{i}, \hat{R}_{\theta}\mathbf{j}}=e^{\pm 2i\theta}\Delta_{\mathbf{i},\mathbf{j}}$, $\Delta_{\hat{\sigma} \mathbf{i}, \hat{\sigma} \mathbf{j}} \neq\pm\Delta_{\mathbf{i},\mathbf{j}}$ \\
\hline
\end{tabular}
\end{table*}

A BCS-MF study is performed on Eq.~\eqref{heff} \cite{Supplementary}. Noting that the Cooper pairing can only take place near the Fermi level, we transform the real-space pairing order parameter $\Delta_{\mathbf{ij}}$ into the $m$-space as $\tilde{\Delta}_{mn}$ and maintain those $m/n$-states within a narrow energy shell near the Fermi level. A self-consistent MF equation for $\tilde{\Delta}_{mn}$ is obtained at any temperature, leading to the following linearized equation at $T_c$ \cite{Supplementary},
\begin{equation}\label{linearized_gap_eq}
\sum_{m'n'}F_{mn,m'n'}\tilde{\tilde{\Delta}}_{m'n'}=\tilde{\tilde{\Delta}}_{mn},
\end{equation}
with $\tilde{\tilde{\Delta}}_{mn}=\tilde{\Delta}_{mn}f_{mn}$, where
\begin{equation}
f_{mn}=\sqrt{(n_{F}(-\tilde{\epsilon}_{n})-n_{F}(\tilde{\epsilon}_{m}))/(\tilde{\epsilon}_{m}+\tilde{\epsilon}_{n})}.
\end{equation}
The formula $F_{mn,m'n'}$ for the singlet pairing is given as
\begin{multline}
F^{(s)}_{mn,m'n'}=-f_{mn}f_{m'n'}\bigg[U\sum_{\mathbf{i}}\xi_{\mathbf{i}m}\xi_{\mathbf{i}n}\xi_{\mathbf{i}m'}\xi_{\mathbf{i}n'}\\
+\frac{U^2}{4}\sum_{\mathbf{i},\mathbf{j}}\chi_{\mathbf{ij}}(\xi_{\mathbf{i}m}\xi_{\mathbf{j}n}+\xi_{\mathbf{i}n}\xi_{\mathbf{j}m})\times(m,n \Rightarrow m', n')\bigg],
\label{F-single}
\end{multline}
and for the triplet case it is
\begin{eqnarray}
\nonumber
F^{(t)}_{mn,m'n'}&=&f_{mn}f_{m'n'}\bigg[\frac{U^2}{4}\sum_{\mathbf{i,j}}\chi_{\mathbf{ij}}(\xi_{\mathbf{i}m}\xi_{\mathbf{j}n}\\
&&-\xi_{\mathbf{i}n}\xi_{\mathbf{j}m})\times(m,n \Rightarrow m', n')\bigg].
\label{F-triple}
\end{eqnarray}

The linearized gap equation~\eqref{linearized_gap_eq} takes the form of an eigenvalue problem of the matrix $F_{mn,m'n'}$ (here we take the combined $mn$ or $m'n'$ as one index), wherein its largest eigenvalue attains 1 at $T_c$, with the corresponding eigenvector $\tilde{\tilde{\Delta}}_{mn}$ determining the pairing symmetry. Due to the lack of translation symmetry here, the real-space gap function $\Delta_{\mathbf{i,j}}$ is no longer just a function of $\mathbf{i-j}$, but a binary function of both $\mathbf{i}$ and $\mathbf{j}$. The situation is similar in the $m$-space. The $m,n$-dependence of $|\tilde{\Delta}_{mn}|$ is shown in  Fig.~\ref{fig:Anderson_theorems} for two typical solutions solved from Eq.~\eqref{linearized_gap_eq}, where for each $m$ there is no unique $n$ which makes $|\tilde{\Delta}_{mn}|$ dominate that of any other $n$, distinct from the result for $U<0$ wherein $|\tilde{\Delta}_{mm}|\gg |\tilde{\Delta}_{mn}(n\ne m)|$ \cite{attractive}. Such a behavior breaks the Anderson's theorem applied for the strong-disorder-limit superconductors.

{\bf Pairing symmetries and phase-diagram:} The classification of pairing symmetries here is based on the symmetry of the linearized gap equation~\eqref{linearized_gap_eq} \cite{Supplementary}. It's proved that the ``pairing potential'' $F_{mn,m'n'}$ is invariant under the $D_5$ point group. Consequently, the set of solutions $\{\tilde{\tilde{\Delta}}^{(\alpha)}_{mn}\}$ ($\alpha=1$ or $1,2$) of Eq.~\eqref{linearized_gap_eq} corresponding to the same $T_c$ furnish an IR of $D_5$. This statement also holds for the real-space gap function $\Delta^{(\alpha)}_{\mathbf{i,j}}$\cite{Supplementary}:
\begin{equation}\label{symmetry}
\Delta^{(\alpha)}_{\hat{g}\mathbf{i},\hat{g}\mathbf{j}}=\sum_{\alpha^{\prime}}D^{(g)}_{\alpha\alpha^{\prime}}\Delta^{(\alpha^{\prime})}_{\mathbf{i},\mathbf{j}},
\end{equation}
with any $g\in D_5$. Then, from the IR which the set of matrices $\{D^{(g)}\}$ belong to, one can judge the pairing symmetry of the state with gap function $\Delta^{(\alpha)}_{\mathbf{i,j}}$.

The four IRs of the $D_5$ point group are listed in Table~\ref{tab:classification}, including two 1D IRs, i.e. $A_1$ and $A_2$, and two 2D IRs, i.e. $E_1$ and $E_2$. For each IR, we list the representation matrices for the two generators of $D_5$, i.e. the $C_{2\pi/5}$ and $\sigma_x$, up to an arbitrary unitary transformation. Each IR listed in Table~\ref{tab:classification} corresponds to one pairing-symmetry class. The identity representation $A_1$ is the $s$-wave with angular momentum $l=0$. The $A_2$ representation is the $h_{y^5-10x^2y^3+5x^4y}$-wave with $l=5$ which is $\sigma$-reflection odd. The $E_1$ ($E_2$) representation provides the doubly-degenerate $p$-wave ($d$-wave) with $l=1$ ($l=2$).

Note that for each of the pairing symmetry listed in Table~\ref{tab:classification}, both spin-singlet and spin-triplet pairings are possible, suggesting that the pairing angular momentum $l$ and the spin statistics are independent. Such independence between the former and the latter is general on all QC lattices due to the lack of translation symmetry. Generally, in a singlet (triplet) pairing state where the spin part of the Cooper-pair wave function is exchange- odd (even), the Fermi statistics requires the spacial part to be exchange- even (odd). The exchange operation in the latter case can be viewed as a 180$^o$-rotation about the mass center of the Cooper pair, and thus this exchange parity is related to the angular momentum $\tilde{l}$ of the moving Cooper pair about its mass center. However, without translation symmetry, $\tilde{l}$ is different from $l$, as the latter is with respect to the fixed coordinate origin. Therefore, on QC lattices, the pairing angular momentum $l$ and the spin statistics are unrelated. Note that such independence between the former and latter can also originate from the lack of inversion symmetry, which can also take place on non-centrosymmetric periodic lattices\cite{footnote3}.

\begin{figure}[htbp]
\includegraphics[width=0.45\textwidth]{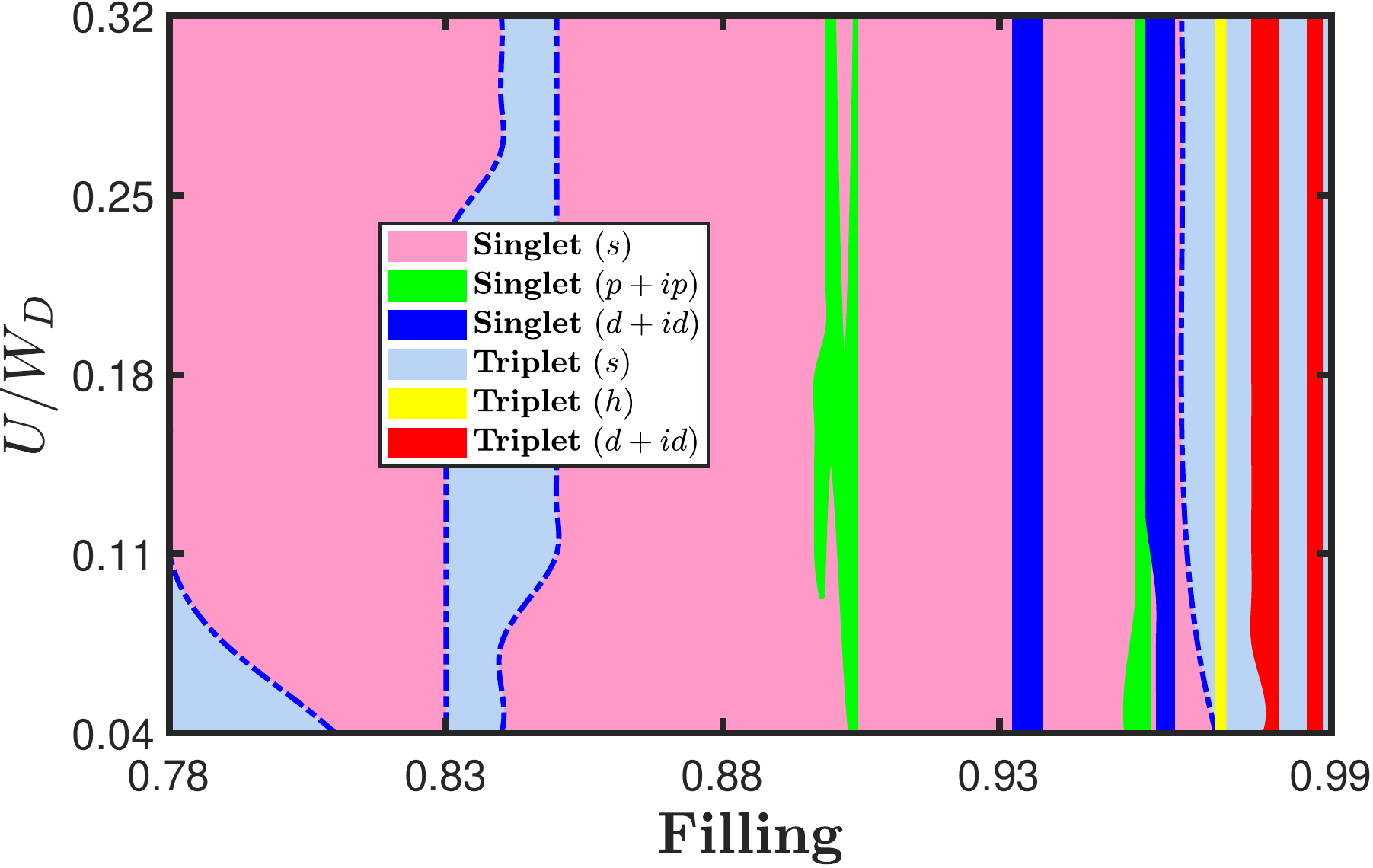}
\caption{\label{fig:phase_diagram}(Color online) Ground-state pairing phase diagram in the filling-interaction plane. The interaction strength of $U$ is limited within a weak-coupling range of $(0,W_{D}/3)$.}
\end{figure}

The pairing phase diagram is shown in Fig.~\ref{fig:phase_diagram} obtained through solving Eq.~\eqref{linearized_gap_eq} for the singlet and triplet channels separately. In our calculations, we adopt a lattice with 13926 sites with open-boundary condition. We focus on the filling range of $(0.78,0.99)$ wherein the DOS is relatively large and the $T_c$ is relatively high. The Hubbard-$U$ adopted here is within a weak-coupling range of $(0,W_{D}/3)$. For the sake of reducing the computation complexity, we limit the states marked by $m^{(\prime)}/n^{(\prime)}$ in Eq.~\eqref{linearized_gap_eq} to Eq.~\eqref{F-triple} within a narrow energy window near the Fermi level containing about 100 states. From Fig.~\ref{fig:phase_diagram}, the obtained pairing symmetries slightly depend on $U/W_D$ but strongly depend on the filling level.  Six out of the eight possible pairing states listed in Table~\ref{tab:classification} are obtained, including the singlet and triplet $s$- and $d$- waves, the singlet $p$-wave and the triplet $h$-wave pairing symmetries.

{\bf Exotic TSCs:} The spin-singlet and spin-triplet $p$- and $d$-wave pairings states listed in Table~\ref{tab:classification} or Fig.~\ref{fig:phase_diagram} belong to 2D IRs of the point group, suggesting the existence of doubly degenerate gap functions $\Delta_{mn}^{(1,2)}$, which would be mixed below $T_c$ to lower the free energy. At $T=0$, the minimization of the expectation value of the effective Hamiltonian~\eqref{eff_hamiltonian} in the BdG MF ground state with gap form factor $\Delta_{mn}^{(1)}+\alpha\Delta_{mn}^{(2)}$ yields $\alpha=\pm i$ for all these cases. Therefore such degenerate doublets would be mixed as $p+ip$ and $d+id$ in the ground state. The real-space gap functions $\Delta_{\mathbf{i,j}}=\Delta_{\mathbf{i,j}}^{(1)}\pm i\Delta_{\mathbf{i,j}}^{(2)}$ of these mixed states show nontrivial winding-number structures: with each $\theta$-angle rotation ($\theta=2n\pi/5$) about the lattice center performed on combined $(\mathbf{i,j})$, the complex phase of $\Delta_{\mathbf{i,j}}$ would be shifted by $\pm l\theta$ ($l$: angular momentum), as listed in Table~\ref{tab:classification}, and shown in Fig.~\ref{fig:Anderson_theorems}(c) and (d) for the $s$- and $d+id$- wave pairings respectively. Such nontrivial winding numbers of these pairing states suggest that they are topologically nontrivial.

\begin{figure}[htbp]
\includegraphics[width=0.4\textwidth]{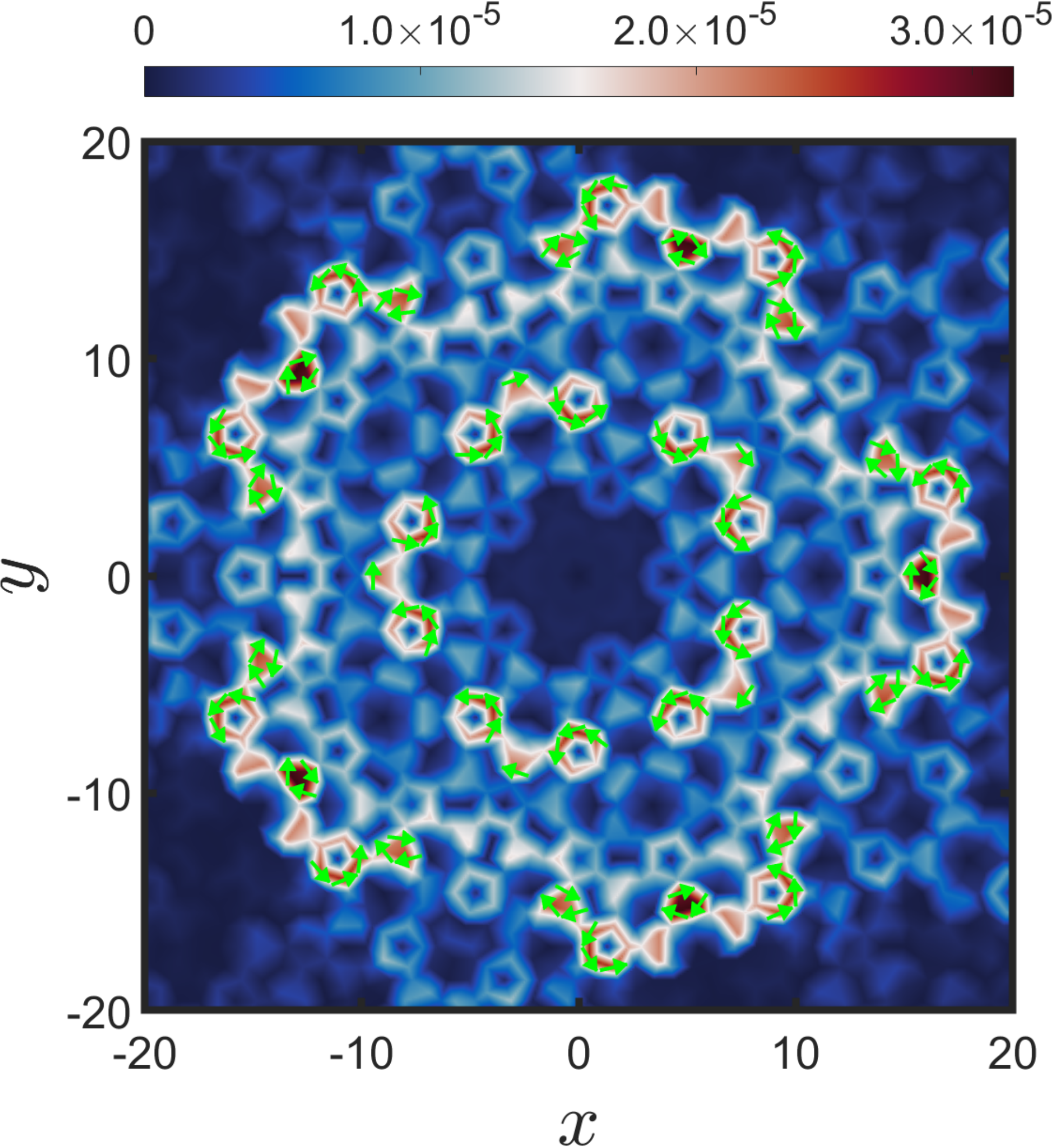}
\caption{\label{fig:spontaneous_flows}(Color online)
Contour plot of the amplitude of the spontaneous bulk super current in the same pairing state as that in Fig. \ref{fig:Anderson_theorems}(b). The green arrows indicate the direction of the current at a specific site. To enhance the visibility, only typical sites are marked with the direction of the current.}
\end{figure}

To better characterize the topology of these pairing states on the QC without translation symmetry \cite{Loring2015,Fulga2016}, we use the {\it K}-theory class characterized by the Chern number. On the finite lattice, based on the spectral-localizer method, the Chern number is obtained as the following pseudo-spectrum invariant index $C_{ps}$ \cite{Supplementary},
\begin{equation}
C_{ps} = \frac{1}{2} Sig
\begin{pmatrix}
  X & Y+iH \\
  Y-iH & -X \\
\end{pmatrix}.
\end{equation}
Here $X$ and $Y$ are the position operators, $H$ is the BdG-Hamiltonian matrix and $Sig$ represents the difference between the numbers of positive and negative eigenvalues of the matrix acted on \cite{Supplementary}. Using this formula, we prove\cite{Supplementary} that any global unitary transformation on the system maintains $C_{ps}$, and that the TR operation changes  the sign of $C_{ps}$, which lead to the following conclusions. Firstly, the $C_{ps}$ of all 1D-IR pairing states are zero. Secondly, for triplet pairing states belonging to the 2D-IRs, the $C_{ps}$ for the TRS-breaking chiral-pairing states $\sum_{\mathbf{ij}}\Delta_{\mathbf{i,j}}(c_{\mathbf{i}\uparrow}c_{\mathbf{j}\downarrow}+c_{\mathbf{i}\downarrow}c_{\mathbf{j}\uparrow})+h.c.$ (or  $\sum_{\mathbf{ij}}\Delta_{\mathbf{i,j}}(c_{\mathbf{i}\uparrow}c_{\mathbf{j}\uparrow}\pm c_{\mathbf{i}\downarrow}c_{\mathbf{j}\downarrow})+h.c.$) are twice of those for the spinless system with $\sum_{\mathbf{ij}}\Delta_{\mathbf{i,j}}c_{\mathbf{i}}c_{\mathbf{j}}+h.c.$, and those for the TRI helical-pairing states $\sum_{\mathbf{ij}}(\Delta_{\mathbf{i,j}}c_{\mathbf{i}\uparrow}c_{\mathbf{j}\uparrow}\pm \Delta^{*}_{\mathbf{i,j}}c_{\mathbf{i}\downarrow}c_{\mathbf{j}\downarrow})+h.c.$ are zero. These triplet pairing states are degenerate here without considering the spin-orbit coupling. Our numerical calculations on the 2D-IR singlet and chiral-triplet pairing states appearing in the phase diagram yield that their Chern numbers are generally integer multiples of twice of their spacial angular momenta, suggesting the presence of TSCs without translation symmetry.

A general and remarkable property of the TSCs on a QC is the presence of spontaneous bulk super current caused by the lack of translation symmetry. To illustrate this point, we have calculated the expectation values of the site-dependent current operator $\hat{\mathbf{J}}_{\mathbf{i}}=-\delta{\hat{\mathcal{H}}}/\delta{\mathbf{A}_{\mathbf{i}}}|_{\mathbf{A}=0}=
\frac{i}{2}\sum_{\mathbf{j}\sigma}t_{\mathbf{ij}}(\mathbf{r_j}-\mathbf{r_i})c_{\mathbf{i}\sigma}^{\dagger}c_{\mathbf{j}\sigma}+h.c.$ (see \cite{Supplementary}). Note that $\hat{\mathbf{J}}_{\mathbf{i}}$ is TR odd, whose expectation value $\left\langle\hat{\mathbf{J}}_{\mathbf{i}}\right\rangle$ should vanish in TRI states. However, in the TRS-breaking chiral pairing states belonging to the 2D-IR, our numerical results shown in Fig. {\ref{fig:spontaneous_flows}} illustrate a five-folded-symmetric pattern with $\left\langle\hat{\mathbf{J}}_{\mathbf{i}}\right\rangle\ne 0$ for any typical site. It's intriguing that the super current forms spontaneous vortices here and there, leading to bulk orbital magnetization that can be detected by experiments. Note that on periodic lattices, the spontaneous super current for a topological superconducting state usually appears at the edge\cite{Horovitz2003,Stone2004,Wang2005,Sauls2011,Kallin2012,Huang2014,Liu2004,Kirtley2007, Curran2014}, although it can also appear in the bulk on complex enough lattices. In the latter case, the averaged current within a unit cell should vanish, otherwise the superfluid density (see below) would be infinity. However, on the QCs with no translation symmetry and hence no unit cell, the distribution pattern of the super current is not limited by such a constraint.

{\bf Discussion and Conclusion:} One might wonder whether the lack of translation symmetry on the Penrose lattice would destroy the phase coherence of the pairing state. This puzzle can be settled by investigating the superfluid density $\rho_s$ defined as $\rho^{\alpha\beta}_s\equiv\lim_{\mathbf{A}\to 0}\frac{-\langle\mathbf{A}|\hat{J}_{\alpha}[\mathbf{A}]|\mathbf{A}\rangle}{A_{\beta}}$, with $\alpha/\beta=x,y$ \cite{Supplementary}. Here a weak uniform vector potential $\mathbf{A}$ along the $\beta$- direction is coupled with the system, $\hat{\mathbf{J}}[\mathbf{A}]=-\delta{\hat{\mathcal{H}}[\mathbf{A}]}/\delta{\mathbf{A}}$ and $|\mathbf{A}\rangle$ represents the ground state of $\hat{\mathcal{H}}[\mathbf{A}]$. It's proved here \cite{Supplementary} that $\rho^{\alpha\beta}_s=\rho_0\delta_{\alpha\beta}$ and our numerical result yields $\rho_0>0$, suggesting a true superconducting state with nonzero superfluid density and hence measurable Meissner effect.

The real-space perturbative approach engaged here and the insight acquired from this work would also apply to other QCs. Particularly, the recently synthesized 30$^{\circ}$- twisted bilayer graphene \cite{TBG1,TBG2} provides a relevant platform for the QC Hubbard model studied here. Similar exotic TSCs would be detected there with proper doping. More interestingly, the $D_{12}$ point group of that QC system leads to more IRs than those of the Penrose lattice. Consequently, exotic TSCs with winding numbers of 3, 4 and 5 are possible, higher than the 1 ($p+ip$) or 2 ($d+id$) obtained here or in periodic systems.

In conclusion, we have performed a real-space perturbative calculation for the Hubbard model on the Penrose lattice. Our results reveal various classes of unconventional SCs induced via the Kohn-Luttinger mechanism. We have classified the pairing symmetries according to the IRs of the $D_5$ point group of this lattice, with most of them exhibited in the pairing phase diagram. Remarkably, each pairing symmetry can be both spin-singlet and spin-triplet. All the 2D-IR pairing states can be TRS-breaking chiral TSCs hosting spontaneous bulk super current and spontaneous vortices. These pairing-mechanism-independent exotic properties of the SCs on the Penrose lattice are caused by the combination of the point-group symmetry and the lack of translation symmetry, and are thus general for all QC lattices, and are rare on periodic lattices. Our work starts the new area of unconventional SCs driven by repulsive interactions on the QC.

\section*{Acknowledgements}
We acknowledge stimulating discussions with Wen Huang. This work is supported by the NSFC (Grant Nos. 11674025, 11704029, 11922401).

\renewcommand{\theequation}{A\arabic{equation}}
\setcounter{equation}{0}
\renewcommand{\thefigure}{A\arabic{figure}}
\setcounter{figure}{0}
\renewcommand{\thetable}{A\arabic{table}}
\setcounter{table}{0}
\begin{widetext}
\section{\label{sec:level1}The real-space perturbative theory}
Let's start from the following positive-U Hubbard model on the Penrose lattice,
\begin{eqnarray}
\mathcal{H} & = & -\sum_{\text{\ensuremath{\mathbf{i}}}\mathbf{j}\sigma}t_{\mathbf{i}\mathbf{j}}c_{\mathbf{i}\sigma}^{\dagger}c_{\mathbf{j}\sigma}+
U\sum_{\mathbf{i}}n_{\mathbf{i}\uparrow}n_{\mathbf{i}\downarrow}-\mu\sum_{\mathbf{i}\sigma}n_{\mathbf{i}\sigma}.\label{eq:original-Hamiltonian}
\end{eqnarray}
To treat with this interacting system, let's first investigate the tight-binding (TB) model in its kinetic-energy part, which can be diagonalized as
\begin{eqnarray}\label{Hubbard}
\mathcal{H}_{\text{TB}}  =  -\sum_{\text{\ensuremath{\mathbf{i}}}\mathbf{j}\sigma}t_{\mathbf{i}\mathbf{j}}c_{\mathbf{i}\sigma}^{\dagger}c_{\mathbf{j}\sigma}-\mu\sum_{\mathbf{i}\sigma}n_{\mathbf{i}\sigma}=\sum_{m}(\epsilon_m-\mu)c^{\dagger}_{m\sigma}c_{m\sigma}\equiv\sum_{m}\tilde{\epsilon}_m c^{\dagger}_{m\sigma}c_{m\sigma}\label{Kinetics}
\end{eqnarray}
Here the index $m$ labels the eigen state on the Penrose lattice, and $c_{m \sigma}=\sum_{\mathbf{i}}\xi_{\mathbf{i}m}c_{\mathbf{i}\sigma}$, with $\xi_{\mathbf{i}m}$ representing for the wave function for the state $m$. In the following, we shall perform a real-space perturbative treatment on the Hubbard interaction.

The Matsubara single-particle Green's funciton is defined as,
\begin{eqnarray}
\mathcal{G}_{\mathbf{i}\mathbf{j}\sigma\sigma'}(\tau_{1},\tau_{2}) = -\langle T_{\tau}c_{\mathbf{i}\sigma}(\tau_{1})c_{\mathbf{j}\sigma'}^{\dagger}(\tau_{2})\rangle=\mathcal{G}_{\mathbf{i}\mathbf{j}\sigma\sigma'}(\tau_{1}-\tau_{2}),\label{eq:Green's-function}
\end{eqnarray}
with $c_{\mathbf{i}\sigma}(\tau)\equiv e^{\mathcal{H}\tau}c_{\mathbf{i}\sigma}e^{-\mathcal{H}\tau}$, and $\langle\cdots\rangle\equiv\mathrm{Tr}[\rho\cdots]$ denotes the thermal average
at the temperature $T$ with $\beta=1/(k_{B}T)$. This Green's function can be Fourier transformed to the imaginary-frequency space as
\begin{equation}
\mathcal{G}_{\mathbf{ij}\sigma\sigma'}(i\omega_{n})\equiv \int ^{\beta}_0 e^{i\omega_n\tau}\mathcal{G}_{\mathbf{ij}\sigma\sigma'}(\tau)d\tau
\end{equation}
In the case of $U=0$, we obtain the bare single-particle Green's function in the eigen basis as
\begin{eqnarray}
\mathcal{G}_{ml\sigma\sigma'}^{(0)}(i\omega_{n}) & = & \frac{\delta_{ml}\delta_{\sigma\sigma'}}{i\omega_{n}-\tilde{\epsilon}_{m}},\label{eq:free-field-Green's-function-state-space}
\end{eqnarray}
which is Fourier transformed to the real space as
\begin{eqnarray}
\mathcal{G}_{\mathbf{ij}\sigma\sigma'}^{(0)}(i\omega_{n}) & = & \sum_{m}\frac{\xi_{\mathbf{i}m}\xi_{\mathbf{j}m}\delta_{\sigma\sigma'}}{i\omega_{n}-\tilde{\epsilon}_{m}},\label{eq:free-field-Green's-function-real-space}
\end{eqnarray}
where $\omega_{n}=(2n+1)\pi/\beta$ is the fermion frequency.

\begin{figure}[htbp]
\includegraphics[width=0.48\textwidth]{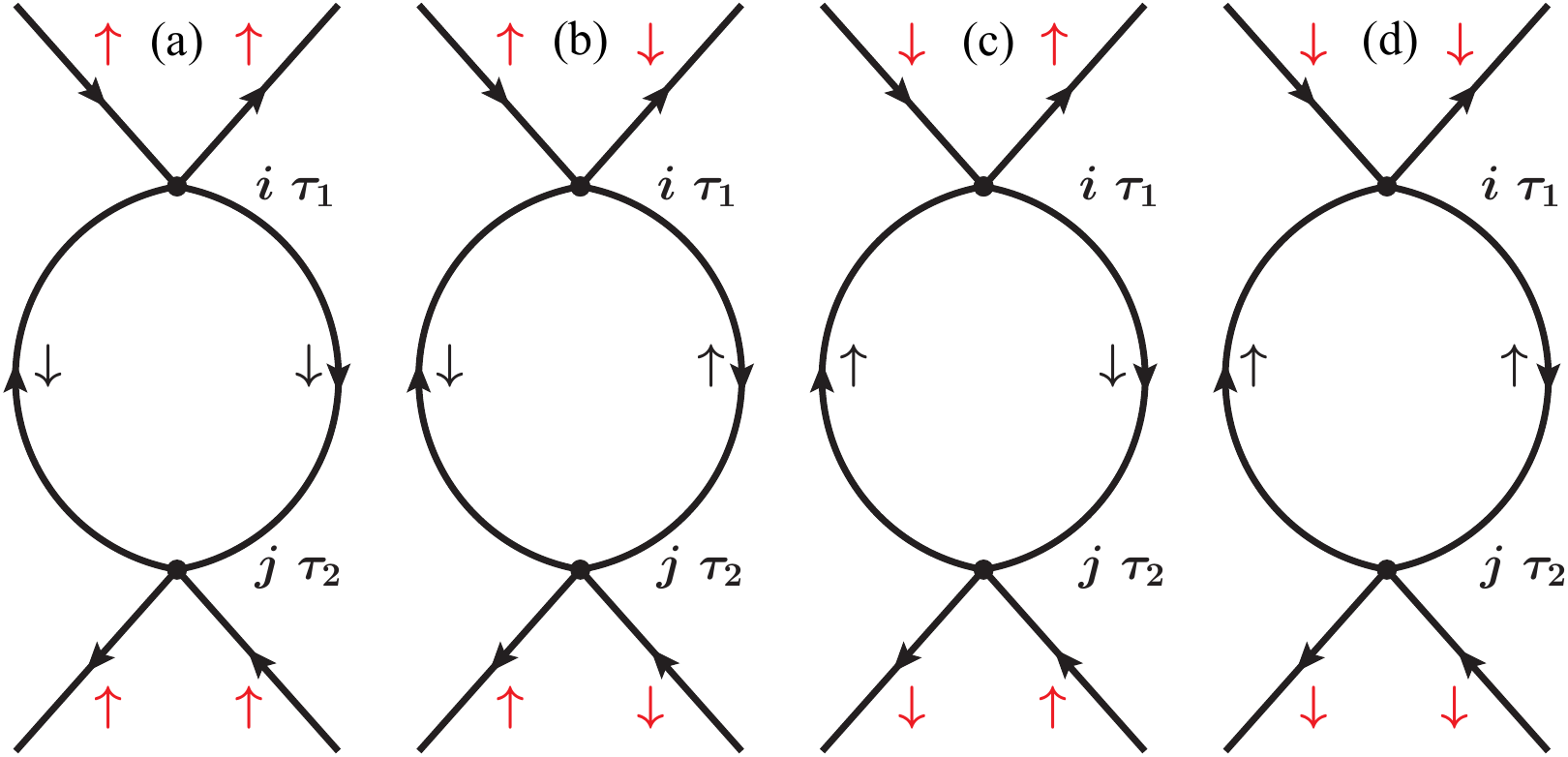}
\caption{\label{fig:Feynman-diagrams}(Color online) Feynman diagrams for the
four second-order perturbative processes of exchanging particlehole
fluctuations.}
\end{figure}

Let's define the real-space susceptibility function
\begin{equation}
\chi_{\mathbf{i}\mathbf{j}}(\tau)\equiv\langle T_{\tau}c_{\mathbf{i}\sigma}^{\dagger}(\tau)c_{\mathbf{i}\sigma^{\prime}}(\tau)c_{\mathbf{j\sigma^{\prime}}}^{\dagger}c_{\mathbf{j}\sigma}\rangle_{0,c}=\langle T_{\tau}c_{\mathbf{i}\uparrow}^{\dagger}(\tau)c_{\mathbf{i}\uparrow}(\tau)c_{\mathbf{j\uparrow}}^{\dagger}c_{\mathbf{j}\uparrow}\rangle_{0,c}.\label{eq:bare-susceptibililty-time}
\end{equation}
Here we only consider the connected Feyman's diagrams in the bare level. Employing Wick's theorem and Eq. (\ref{eq:free-field-Green's-function-real-space}), Eq. (\ref{eq:bare-susceptibililty-time}) can be evatuated, whose Fourier transformation to the imaginary-frequency space is given as
\begin{eqnarray}
\chi_{\mathbf{ij}}(i\Omega_{n}) & = & \sum_{ml}\xi_{\mathbf{i}m}\xi_{\mathbf{j}m}\xi_{\mathbf{i}l}\xi_{\mathbf{j}l}\frac{n_{F}(\tilde{\epsilon}_{m})-n_{F}(\tilde{\epsilon}_{l})}{i\Omega_{n}+\tilde{\epsilon}_{l}-\tilde{\epsilon}_{m}},\label{eq:bare-susceptibility-frequency}
\end{eqnarray}
where $\Omega_{n}=2n\pi/\beta$ is the boson frequency.

In the Kohn-Luttinger mechanism\cite{KL1,KL2}, two electrons at the Fermi level acquire an effective interaction through exchanging the particle-hole fluctuations. There are four relevant second-order processes at the bare-susceptibility level for this mechanism, which are described by the four Feyman's diagrams shown in Fig. \ref{fig:Feynman-diagrams}, leading to the following effective Hamiltonian,
\begin{eqnarray}\label{effective_Hamiltonian}
\mathcal{H}_{eff} & = & -\sum_{\mathbf{ij}\sigma}t_{\mathbf{ij}}c_{\mathbf{i}\sigma}^{\dagger}c_{\mathbf{j}\sigma}+U\sum_{\mathbf{i}}n_{\mathbf{i}\uparrow}n_{\mathbf{i}\downarrow}-
\mu\sum_{\mathbf{i}\sigma}n_{\mathbf{i}\sigma}-(U^{2}/2)\sum_{\mathbf{ij}\sigma\sigma'}\chi_{\mathbf{ij}}c_{\mathbf{i}\sigma}^{\dagger}c_{\mathbf{i}\sigma'}
c_{\mathbf{j}\sigma'}^{\dagger}c_{\mathbf{j}\sigma},
\end{eqnarray}
with $\chi_{\mathbf{ij}}\equiv\chi_{\mathbf{ij}}(i\Omega_{n}=0)$.

The role of the effective Hamiltonian (\ref{effective_Hamiltonian}) lies in that, for the calculation of the abnormal Green's function representing the pairing order parameter, the result obtained at the mean-field (MF) level using Hamiltonian (\ref{effective_Hamiltonian}) can approximately provide the result obtained up to the second-order perturbative expansion of the S-matrix (see Mahan's book \cite{Mahan}, page 87-89) using the original Hamiltonian (\ref{eq:original-Hamiltonian}). Therefore, we are justified to simply perform a MF calculation on the obtained effective Hamiltonian (\ref{effective_Hamiltonian}) to obtain the pairing order parameter, which approximately has the same effect of doing a second-order perturbative calculation on the original Hamiltonian (\ref{eq:original-Hamiltonian}), engaging complicated Green's function skills. Note that the simple MF calculation on Eq. (\ref{eq:original-Hamiltonian}) will not lead to SC, as the interaction here is repulsive. In the following, we shall perform a MF study on Eq. (\ref{effective_Hamiltonian}).

As the effective interaction (\ref{effective_Hamiltonian}) is invariant under spin-SU(2) transformation, we can perform the MF decoupling in the singlet and triplet channels separately. In the singlet channel Eq. (\ref{effective_Hamiltonian}) is MF decoupled as
\begin{eqnarray}
\mathcal{H}_{mf}^{s} & = & \sum_{m\sigma}\tilde{\epsilon}_{m}c_{m\sigma}^{\dagger}c_{m\sigma}+U(\sum_{\mathbf{i}}\Delta_{{\bf \mathbf{i}}}^{s\dagger}\langle\Delta_{\mathbf{i}}^{s}\rangle+h.c.-|\langle\Delta_{\mathbf{i}}^{s}\rangle|^{2})+\frac{U^{2}}{2}(\Delta_{\mathbf{ij}}^{s\dagger}\langle\Delta_{\mathbf{ij}}^{s}\rangle+h.c.-|\langle\Delta_{\mathbf{ij}}^{s}\rangle|^{2})\chi_{\mathbf{ij}}\nonumber \\
 & \equiv & \sum_{m,\sigma}\tilde{\epsilon}_{m}c_{m\sigma}^{\dagger}c_{m\sigma}+\sum_{m, n}\tilde{\Delta}_{mn}^{s}c_{m\uparrow}^{\dagger}c_{n\downarrow}^{\dagger}+h.c.,\label{eq:mean-field-Hamiltonian-singlet}
\end{eqnarray}
with
\begin{eqnarray}
\Delta_{\mathbf{i}}^{s\dagger} & = & c_{\mathbf{i}\uparrow}^{\dagger}c_{\mathbf{i}\downarrow}^{\dagger},\label{eq:pairing-singlet-00-diagonal}\\
\Delta_{\mathbf{ij}}^{s\dagger} & = & \frac{1}{\sqrt{2}}(c_{\mathbf{i}\uparrow}^{\dagger}c_{\mathbf{j}\downarrow}^{\dagger}-
c_{\mathbf{i}\downarrow}^{\dagger}c_{\mathbf{j}\uparrow}^{\dagger}),\label{eq:pairing-singlet-00-off-diagonal}
\end{eqnarray}
The MF decoupling of Eq. (\ref{effective_Hamiltonian}) in the triplet channel can be performed in three channels with the following three order parameters,
\begin{eqnarray}
\Delta_{\mathbf{ij}}^{t\dagger}(1,1) & = & c_{\mathbf{i}\uparrow}^{\dagger}c_{\mathbf{j}\uparrow}^{\dagger},\label{eq:pairing-triplet-11}\\
\Delta_{\mathbf{ij}}^{t\dagger}(1,0) & = & \frac{1}{\sqrt{2}}(c_{\mathbf{i}\uparrow}^{\dagger}c_{\mathbf{j}\downarrow}^{\dagger}+c_{\mathbf{i}\downarrow}^{\dagger}c_{\mathbf{j}\uparrow}^{\dagger}),\label{eq:pairing-triplet-10}\\
\Delta_{\mathbf{ij}}^{t\dagger}(1,-1) & = & c_{\mathbf{i}\downarrow}^{\dagger}c_{\mathbf{j}\downarrow}^{\dagger},\label{eq:pairing-triplet-1-1}
\end{eqnarray}
which represents for the triplet-pairing components with the total $S_z$ of the Cooper pair to be $\hbar,0$ and $-\hbar$ respectively. Due to the spin-SU(2) symmetry, these three channels are exactly degenerate, which allows that we can only study one component, e.g. the $(1,0)$ component, among the three ones. The MF decoupling in this channel reads as
\begin{eqnarray}
\mathcal{H}_{mf}^{t} & = & \sum_{m\sigma}\tilde{\epsilon}_{m}c_{m\sigma}^{\dagger}c_{m\sigma}-\frac{U^{2}}{2}(\Delta_{\mathbf{ij}}^{t\dagger}\langle\Delta_{\mathbf{ij}}^{t}\rangle+h.c.-|\langle\Delta_{\mathbf{ij}}^{t}\rangle|^{2})\chi_{ij}\nonumber \\
 & \equiv & \sum_{m,\sigma}\tilde{\epsilon}_{m}c_{m\sigma}^{\dagger}c_{m\sigma}+\sum_{m,n}\tilde{\Delta}_{mn}^{t}c_{m\uparrow}^{\dagger}c_{n\downarrow}^{\dagger}+h.c.,\label{eq:mean-field-Hamiltonian-triplet}
\end{eqnarray}
with $\Delta^{t\dagger}$ to be the abbriviation of the $\Delta^{t\dagger}$(1,0) in Eq. (\ref{eq:pairing-triplet-10}). To make the calculation feasible, we constrain the summation of energy
within a small window, $\Delta E$, around the chemical potential $\mu$.

The BdG Hamiltonians for both the singlet and the triplet $(1,0)$ channels can be written as,
\begin{eqnarray}
\mathcal{H}_{BdG} = \left(\begin{array}{cc}
c_{\uparrow}^{\dagger} & c_{\downarrow}\end{array}\right)\left(\begin{array}{cc}
\tilde{\epsilon} & \tilde{\Delta}\\
\text{\ensuremath{\tilde{\Delta}^{\dagger}}} & -\tilde{\epsilon}
\end{array}\right)\left(\begin{array}{c}
c_{\uparrow}\\
c_{\downarrow}^{\dagger}
\end{array}\right) \equiv  X^{\dagger}H_{BdG}X,\label{H_BdG}
\end{eqnarray}
where $c_{\sigma}$ is an abbreviation of $\left(c_{1\sigma},\cdots,c_{L\sigma}\right)$
and $L$ is the number of energy levels in the truncated energy window.
The BdG Hamiltonian can be diagonalized as

\begin{eqnarray}
\mathcal{H}_{BdG} & = & \sum_{m=1}^{2L}E_{m}\gamma_{m}^{\dagger}\gamma_{m},
\end{eqnarray}
with $X=\omega\gamma$, where $\omega$ is the matrix of eigenvectors of $H_{BdG}$. In the following, we derive the linearized gap equation at the critical temperature $T_c$, solving which we can obtain $T_c$ and the pairing symmetry.  We shall demonstrate the procedure of deriving the gap equation for the singlet channel as an example in the following, while for the triplet $(1,0)$ channel, we shall only give the results.

The $\tilde{\Delta}$ in Eq. (\ref{H_BdG}) can be self-consistently calculated by expanding $c_{\mathbf{i}\sigma}$ by $c_{m\sigma}$ and further by $\gamma_{m}$.  We demonstrate the derivations as
the following. For the singlet channel at any finite temperature $T$
\begin{eqnarray}
\tilde{\Delta}_{mn}^{s} & = & U\sum_{\mathbf{i}}\xi_{\mathbf{i}m}\xi_{\mathbf{i}n}\langle\Delta_{i}\rangle+\frac{U^{2}}{2\sqrt{2}}\sum_{\mathbf{ij}}\chi_{\mathbf{ij}}(\xi_{\mathbf{i}m}\xi_{\mathbf{j}n}+\xi_{\mathbf{i}n}\xi_{\mathbf{j}m})\langle\Delta_{\mathbf{ij}}\rangle\nonumber \\
 & = & U\sum_{\mathbf{i}}\xi_{\mathbf{i}m}\xi_{\mathbf{i}n}\sum_{\substack{m'n'=1\cdots L\\
m''=1\cdots2L
}
}\omega_{m'+L,m''}^{*}\omega_{n',m''}\xi_{\mathbf{i}m'}\xi_{\mathbf{i}n'}n_{F}(E_{m''})\nonumber \\
 &  & +\frac{U^{2}}{4}\sum_{\mathbf{ij}}\chi_{\mathbf{ij}}(\text{\ensuremath{\xi_{\mathbf{i}m}\xi_{\mathbf{j}n}+\xi_{\mathbf{i}n}\xi_{\mathbf{j}m}}})\sum_{\substack{m'n'=1\cdots L\\
m''=1\cdots2L
}
}\omega_{m'+L,m''}^{*}\omega_{n',m''}(\xi_{\mathbf{j}m'}\xi_{\mathbf{i}n'}+\xi_{\mathbf{j}n'}\xi_{\mathbf{i}m'})n_{F}(E_{m''}).\label{eq:self-single}
\end{eqnarray}
Meanwhile, for triplet,
\begin{eqnarray}
\tilde{\Delta}_{mn}^{t} & = & \frac{-U^{2}}{2\sqrt{2}}\sum_{\mathbf{ij}}\chi_{\mathbf{ij}}(\xi_{\mathbf{i}m}\xi_{\mathbf{j}n}-\xi_{\mathbf{i}n}\xi_{\mathbf{j}m})\langle\Delta_{\mathbf{ij}}\rangle\nonumber \\
 & = & -\frac{U^{2}}{4}\sum_{\mathbf{ij}}\chi_{\mathbf{ij}}(\text{\ensuremath{\xi_{\mathbf{i}m}\xi_{\mathbf{j}n}-\xi_{\mathbf{i}n}\xi_{\mathbf{j}m}}})\sum_{\substack{m'n'=1\cdots L\\
m''=1\cdots2L
}
}\omega_{m'+L,m''}^{*}\omega_{n',m''}(\xi_{\mathbf{j}m'}\xi_{\mathbf{i}n'}-\xi_{\mathbf{j}n'}\xi_{\mathbf{i}m'})n_{F}(E_{m''})\nonumber \\
 & = & \text{\ensuremath{\frac{U^{2}}{4}\sum_{\mathbf{ij}}\chi_{\mathbf{ij}}}(\text{\ensuremath{\xi_{\mathbf{i}m}\xi_{\mathbf{j}n}-\xi_{\mathbf{i}n}\xi_{\mathbf{j}m}}})\ensuremath{\sum_{\substack{m'n'=1\cdots L\\
 m''=1\cdots2L
}
 }\omega_{m'+L,m''}^{*}\omega_{n',m''}}(\ensuremath{\xi_{\mathbf{i}m'}\xi_{\mathbf{j}n'}-\xi_{\mathbf{i}n'}\xi_{\mathbf{j}m'}})\ensuremath{n_{F}}(\ensuremath{E_{m''}}).}\label{eq:self-triple}
\end{eqnarray}
Here $n_{F}(...)$ represents the Fermi-Dirac distribution function. Noting that $E$ and $\omega$ are implicit functions of $\{\tilde{\Delta}^{s}_{mn}\}$, the Eq. (\ref{eq:self-single}) is actually a self-consistent equation. In practice, it contain too many variables to optimize the solution of the lowest free energy easily. Alternatively, we consider the temperature just below $T_{c}$, where the order parameters are infinitely small and we can treat them perturbatively, as done in the following.

The BdG Hamiltonian is made up with two terms,
\begin{eqnarray}
H_{BdG} & = & \left(\begin{array}{cc}
\tilde{\epsilon}\\
 & -\tilde{\epsilon}
\end{array}\right)+\left(\begin{array}{cc}
 & \tilde{\Delta}\\
\text{\ensuremath{\tilde{\Delta}^{\dagger}}}
\end{array}\right).
\end{eqnarray}
When the second term goes to infinitesimal just below $T_c$, we can take it as perturbation, and calculate the $E$ and $\omega$
using perturbation theory. Up to the second-order perturbation, we have
\begin{eqnarray}
E_{m} & = & \left\{ \begin{array}{c}
\tilde{\epsilon}_{m}+\sum_{n\neq m}\frac{|\tilde{\Delta}_{mn}|^{2}}{\tilde{\epsilon}_{m}-\tilde{\epsilon}_{n}}\approx\tilde{\epsilon}_{m},m\leq L,\\
\\
-\tilde{\epsilon}_{m-L}+\sum_{n\neq m}\frac{|\tilde{\Delta}_{mn}|^{2}}{\tilde{\epsilon}_{m}-\tilde{\epsilon}_{n}}\approx-\tilde{\epsilon}_{m-L},m>L,
\end{array}\right.
\end{eqnarray}

\begin{eqnarray}
\omega_{m,n} & = & \delta_{mn},\ \ m,n\leq L,\\
\omega_{m+L,n} & = & \frac{\tilde{\Delta}_{nm}^{*}}{\tilde{\epsilon}_{n}+\tilde{\epsilon}_{m}},\ \ m,n\leq L,\\
\omega_{m,n+L} & = & \frac{\tilde{\Delta}_{mn}}{-\tilde{\epsilon}_{n}-\tilde{\epsilon}_{m}},\ \ m,n\leq L,\\
\omega_{m+L,n+L} & = & \delta_{mn},\ \ m,n\leq L.
\end{eqnarray}
Substituting the perturbative results into Eq. (\ref{eq:self-single})
and Eq. (\ref{eq:self-triple}), and keeping to first order on both
sides, we obtain the linearied gap equaitons,
\begin{align}
\tilde{\Delta}_{mn}^{s} & =\sum_{m'n'}\frac{[n_{F}(\tilde{\epsilon}_{n'})-n_{F}(-\tilde{\epsilon}_{m'})]}{\tilde{\epsilon}_{m'}+\tilde{\epsilon}_{n'}}\Bigg[U\sum_{\mathbf{i}}\xi_{\mathbf{i}n}\xi_{\mathbf{i}m}\xi_{\mathbf{i}n'}\xi_{\mathbf{i}m'}+\frac{U^{2}}{4}\sum_{\mathbf{ij}}\chi_{\mathbf{ij}}(\text{\ensuremath{\xi_{\mathbf{i}m}\xi_{\mathbf{j}n}+\xi_{\mathbf{i}n}\xi_{\mathbf{j}m}}})\times(m,n\Rightarrow m'n')\Bigg]\tilde{\Delta}_{n'm'}^{s}\nonumber \\
 & =-\sum_{m'n'}\frac{[n_{F}(-\tilde{\epsilon}_{n'})-n_{F}(\tilde{\epsilon}_{m'})]}{\tilde{\epsilon}_{m'}+\tilde{\epsilon}_{n'}}\Bigg[U\sum_{\mathbf{i}}\xi_{\mathbf{i}n}\xi_{\mathbf{i}m}\xi_{\mathbf{i}n'}\xi_{\mathbf{i}m'}+\frac{U^{2}}{4}\sum_{\mathbf{ij}}\chi_{\mathbf{ij}}(\text{\ensuremath{\xi_{\mathbf{i}m}\xi_{\mathbf{j}n}+\xi_{\mathbf{i}n}\xi_{\mathbf{j}m}}})\times(m,n\Rightarrow m'n')\Bigg]\tilde{\Delta}_{m'n'}^{s},\label{eq:gap-equation-singlet}
\end{align}
and
\begin{align}
\tilde{\Delta}_{mn}^{t} & =\sum_{m'n'}\frac{[n_{F}(\tilde{\epsilon}_{n'})-n_{F}(-\tilde{\epsilon}_{m'})]}{\tilde{\epsilon}_{m'}+\tilde{\epsilon}_{n'}}\Bigg[\frac{U^{2}}{4}\sum_{\mathbf{ij}}\chi_{\mathbf{ij}}(\text{\ensuremath{\xi_{\mathbf{i}m}\xi_{\mathbf{j}n}-\xi_{\mathbf{i}n}\xi_{\mathbf{j}m}}})\times(m,n\Rightarrow m'n')\Bigg]\tilde{\Delta}_{n'm'}^{t}\nonumber \\
= & \sum_{m'n'}\frac{[n_{F}(-\tilde{\epsilon}_{n'})-n_{F}(-\tilde{\epsilon}_{m'})]}{\tilde{\epsilon}_{m'}+\tilde{\epsilon}_{n'}}\Bigg[\frac{U^{2}}{4}\sum_{\mathbf{ij}}\chi_{\mathbf{ij}}(\text{\ensuremath{\xi_{\mathbf{i}m}\xi_{\mathbf{j}n}-\xi_{\mathbf{i}n}\xi_{\mathbf{j}m}}})\times(m,n\Rightarrow m'n')\Bigg]\tilde{\Delta}_{m'n'}^{t},\label{eq:gap-equation-triplet}
\end{align}
which are denoted as

\begin{eqnarray}
\tilde{\Delta}_{mn}^{s/t} & = & F_{mn,m'n'}^{(s/t)}\tilde{\Delta}_{m'n'}^{s/t}.\label{eq:linearized}
\end{eqnarray}

The problem defined in Eq. (\ref{eq:linearized}) now becomes to seek the eigenvector(s) of the matrix $F^{(s/t)}$ (here we have taken the $mn$ as the row index and the $m^{\prime}n^{\prime}$ as the column index ) with unit eigenvalue. It is noticed that the $F^{(s/t)}$ is not Hermition, so we turn to solve another equivalent eigenvalue problem for an Hermition matrix. Let's define
\begin{equation}\label{redefine}
\tilde{\tilde{\Delta}}^{s/t}_{mn}\equiv\tilde{\Delta}^{s/t}_{mn}\sqrt{\frac{[n_{F}(-\tilde{\epsilon}_{n})-n_{F}(\tilde{\epsilon}_{m})]}{\tilde{\epsilon}_{m}+\tilde{\epsilon}_{n}}},
\end{equation}
then the Eq. (\ref{eq:linearized}) becomes
\begin{eqnarray}
\tilde{\tilde{\Delta}}_{mn}^{s/t} & = & \tilde{F}_{mn,m'n'}^{(s/t)}\tilde{\tilde{\Delta}}_{m'n'}^{s/t},\label{eq:linearized-symmetrized}
\end{eqnarray}
 where $\tilde{F}^{(s)}$ and $\tilde{F}^{(t)}$ are given by
\begin{eqnarray}
\tilde{F}_{mn,m'n'}^{(s)} & = & -\sqrt{\frac{[n_{F}(-\tilde{\epsilon}_{n})-n_{F}(\tilde{\epsilon}_{m})]}{\tilde{\epsilon}_{m}+\tilde{\epsilon}_{n}}}\Bigg[U\sum_{\mathbf{i}}\xi_{\mathbf{i}n}\xi_{\mathbf{i}m}\xi_{\mathbf{i}n'}\xi_{\mathbf{i}m'}\nonumber \\
 &  & +\frac{U^{2}}{4}\sum_{\mathbf{ij}}\chi_{\mathbf{ij}}(\text{\ensuremath{\xi_{\mathbf{i}m}\xi_{\mathbf{j}n}+\xi_{\mathbf{i}n}\xi_{\mathbf{j}m}}})\times(m,n\Rightarrow m'n')\Bigg]\sqrt{\frac{[n_{F}(-\tilde{\epsilon}_{n'})-n_{F}(\tilde{\epsilon}_{m'})]}{\tilde{\epsilon}_{m'}+\tilde{\epsilon}_{n'}}}
\end{eqnarray}
and
\begin{eqnarray}
\tilde{F}_{mn,m'n'}^{(t)} & = & \sqrt{\frac{[n_{F}(-\tilde{\epsilon}_{n})-n_{F}(\tilde{\epsilon}_{m})]}{\tilde{\epsilon}_{m}+\tilde{\epsilon}_{n}}}\Bigg[\frac{U^{2}}{4}\sum_{\mathbf{ij}}\chi_{\mathbf{ij}}(\text{\ensuremath{\xi_{\mathbf{i}m}\xi_{\mathbf{j}n}-\xi_{\mathbf{i}n}\xi_{\mathbf{j}m}}})\times(m,n\Rightarrow m'n')\Bigg]\sqrt{\frac{[n_{F}(-\tilde{\epsilon}_{n'})-n_{F}(\tilde{\epsilon}_{m'})]}{\tilde{\epsilon}_{m'}+\tilde{\epsilon}_{n'}}}.
\end{eqnarray}
Clearly, the matrix $\tilde{F}^{(s/t)}$ now becomes real and symmetric, which is therefore Hermition.

The linearized gap equation Eq. (\ref{eq:linearized-symmetrized}) takes the form of an eigenvalue problem for the matrix $\tilde{F}^{(s)}$ or $\tilde{F}^{(t)}$. The superconducting critical temperature $T_c$ is the temperature at which the largest eigenvalue of  $\tilde{F}^{(s)}$ or $\tilde{F}^{(t)}$ attains 1. The pairing symmetry is determined by the relative gap function $\tilde{\Delta}_{mn}^{s}$ or $\tilde{\Delta}_{mn}^{t}$, which is related to the eigenvector $\tilde{\tilde{\Delta}}_{mn}^{s}$ or $\tilde{\tilde{\Delta}}_{mn}^{t}$ corresponding to the largest eigenvalue, i.e. 1, of the matrix $\tilde{F}^{(s)}$ or $\tilde{F}^{(t)}$ through the relation (\ref{redefine}). Note that $\tilde{\Delta}_{mn}^{s}$ or $\tilde{\Delta}_{mn}^{t}$ only serves as a renormalized gap form factor, and the global pairing amplitude will be enhanced with the decreasing of $T$ below $T_c$, determined by the minimization of free energy. Note that in the main text, we have written the $\tilde{F}_{mn,m'n'}^{(s/t)}$ on the above as $F_{mn,m'n'}^{(s/t)}$.

\section{Basis for Classification of Pairing Symmetries}

Here, we prove that if the binary gap function $\Delta_{\mathbf{i},\mathbf{j}}$ is the real-space correspondence to a solution of the linearized gap equation Eq. (\ref{eq:linearized}), then $\Delta_{\hat{g}\mathbf{i},\hat{g}\mathbf{j}}$ will also be a real-space gap function corresponding to the solution of the equation with the same critical temperature. As a result, all the gap functions $\{\Delta^{(\alpha)}_{\mathbf{i},\mathbf{j}}\}$ ($\alpha=1,2,\cdots$) corresponding to the solution of the linearized gap equation with the same critical temperature just form an irreducible representation (IR) of the $D_5$ point group of the system. This builds the basis for the classification of the pairing symmetries.

Considering each element $g\in D_5$, we have $\hat{P}_{g}|\mathbf{i},\sigma\rangle\equiv|\hat{g}\mathbf{i},\sigma\rangle$, from which we obtain
\begin{equation}
\hat{P}_{g}c_{\mathrm{\mathbf{i}}\sigma}\hat{P}_{g}^{-1}=c_{\hat{g}\mathbf{i},\sigma}.
\end{equation}
Then, the effect of $\hat{P}_{g}$ acting on an eigenvector of the TB Hamiltonian is given by
\begin{eqnarray}
\hat{P}_{g}\xi_{\mathbf{i},m} & = & \xi_{\hat{g}^{-1}\mathbf{i},m}.
\end{eqnarray}
On the other hand, as the TB Hamiltonian is invariant under $D_5$, its eigenvector(s) corresponding to the same eigenvalue must furnish an IR of the point group. As a result, we have
\begin{equation}
\hat{P}_{g}\xi_{\mathbf{i},m}=\xi_{\hat{g}^{-1}\mathbf{i},m}\equiv\xi_{\hat{g}^{-1}\mathbf{i},\tilde{m}\tilde{\alpha}}=\sum_{\tilde{\alpha}'}D_{\tilde{\alpha},\tilde{\alpha}'}^{(g,\tilde{m})}\xi_{\mathbf{i},\tilde{m}\tilde{\alpha}'}.\label{eq:representation-on-eigenvector-1-2}
\end{equation}
Here we explicitly express the index $m$ of an eigen state with a pair of indices, i.e. $m\equiv\tilde{m}\tilde{\alpha}$, with the first index labeling the energy level and the second one labeling the degenerate eigen state(s) with the same eigen energy. The $D^{(g,\tilde{m})}$ is the matrix corresponding to the element $g$ for the IR furnished by the eigenvector(s) belonging
to $\tilde{\epsilon}_{\tilde{m}}$. Making use of the obviously equality, $\chi_{\mathbf{i},\mathbf{j}}=\chi_{\hat{g}\mathbf{i},\hat{g}\mathbf{j}}$, it is convenient to show that the $F^{(s)}$ is invariant under the following symmetry-group operation,
\begin{eqnarray}\label{invariant}
&\sum&_{\tilde{\mu}\tilde{\nu}\tilde{\mu}'\tilde{\nu}'}D_{\tilde{\alpha},\tilde{\mu}}^{(g,\tilde{m})}D_{\tilde{\beta},\tilde{\nu}}^{(g,\tilde{n})}D_{\tilde{\alpha}',\tilde{\mu}'}^{(g,\tilde{m}')}D_{\tilde{\beta}',\tilde{\nu}'}^{(g,\tilde{n}')}F_{\tilde{m}\tilde{\mu}\tilde{n}\tilde{\nu},\tilde{m}'\tilde{\mu}'\tilde{n}'\tilde{\nu}'}^{(s)}=\frac{[n_{F}(\tilde{\epsilon}_{\tilde{m}'})-n_{F}(-\tilde{\epsilon}_{\tilde{n}'})]}{\tilde{\epsilon}_{\tilde{m}'}+\tilde{\epsilon}_{\tilde{n}'}}\Bigg[U\sum_{\mathbf{i}}\xi_{\hat{g}^{-1}\mathbf{i},\tilde{n}\tilde{\beta}}\xi_{\hat{g}^{-1}\mathbf{i},\tilde{m}\tilde{\alpha}}\xi_{\hat{g}^{-1}\mathbf{i},\tilde{n}'\tilde{\beta}'}\xi_{\hat{g}^{-1}\mathbf{i},\tilde{m}'\tilde{\alpha}'}
\nonumber\\&&+\frac{U^{2}}{4}\sum_{\mathbf{ij}}\chi_{\hat{g}^{-1}\mathbf{i},\hat{g}^{-1}\mathbf{j}}(\text{\ensuremath{\xi_{\hat{g}^{-1}\mathbf{i},\tilde{m}\tilde{\alpha}}\xi_{\hat{g}^{-1}\mathbf{j},\tilde{n}\tilde{\beta}}+\xi_{\hat{g}^{-1}\mathbf{i},\tilde{n}\tilde{\beta}}\xi_{\hat{g}^{-1}\mathbf{j},\tilde{m}\tilde{\alpha}}}})\times(\tilde{m}\tilde{\alpha},\tilde{n}\tilde{\beta}\Rightarrow\tilde{m}'\tilde{\alpha}',\tilde{n}'\tilde{\beta}')\Bigg]
\nonumber\\&=&\frac{[n_{F}(\tilde{\epsilon}_{\tilde{m}'})-n_{F}(-\tilde{\epsilon}_{\tilde{n}'})]}{\tilde{\epsilon}_{\tilde{m}'}+\tilde{\epsilon}_{\tilde{n}'}}\Bigg[U\sum_{\mathbf{i}}\xi_{\mathbf{i},\tilde{n}\tilde{\beta}}\xi_{\mathbf{i},\tilde{m}\tilde{\alpha}}\xi_{\mathbf{i},\tilde{n}'\tilde{\beta}'}\xi_{\mathbf{i},\tilde{m}'\tilde{\alpha}'}+\frac{U^{2}}{4}\sum_{\mathbf{ij}}\chi_{\mathbf{ij}}(\text{\ensuremath{\xi_{\mathbf{i},\tilde{m}\tilde{\alpha}}\xi_{\mathbf{j},\tilde{n}\tilde{\beta}}+\xi_{\mathbf{i},\tilde{n}\tilde{\beta}}\xi_{\mathbf{j},\tilde{m}\tilde{\alpha}}}}){\color{red}\times}(\tilde{m}\tilde{\alpha},\tilde{n}\tilde{\beta}\Rightarrow\tilde{m}'\tilde{\alpha}',\tilde{n}'\tilde{\beta}')\Bigg]
\nonumber\\&=&F_{\tilde{m}\tilde{\alpha}\tilde{n}\tilde{\beta},\tilde{m}'\tilde{\alpha}'\tilde{n}'\tilde{\beta}'}^{(s)},
\end{eqnarray}
and so as $F^{(t)}$. From Eq. (\ref{invariant}), it can be easily proved that if $\tilde{\Delta}^{s/t}_{mn}\equiv\tilde{\Delta}^{s/t}_{\tilde{m}\tilde{\alpha}\tilde{n}\tilde{\beta}}$ is a solution of Eq. (\ref{eq:gap-equation-singlet}) or Eq. (\ref{eq:gap-equation-triplet}), then $(\hat{P}_{g}\tilde{\Delta}^{s/t})_{\tilde{m}\tilde{\alpha}\tilde{n}\tilde{\beta}}\equiv \sum_{\tilde{\mu}\tilde{\nu}}D_{\tilde{\alpha},\tilde{\mu}}^{(g,\tilde{m})}D_{\tilde{\beta},\tilde{\nu}}^{(g,\tilde{n})}\tilde{\Delta}^{s/t}_{\tilde{m}\tilde{\mu}\tilde{n}\tilde{\nu}}$ would also be a solution of Eq. (\ref{eq:gap-equation-singlet}) or Eq. (\ref{eq:gap-equation-triplet}) corresponding to the same $T_c$. Let $\Delta^{s/t}_{\mathbf{ij}}$ be the real-space correspondence of $\tilde{\Delta}^{s/t}_{mn}$ via the relation $\Delta^{s/t}_{\mathbf{ij}}=\sum_{mn}\tilde{\Delta}^{s/t}_{mn}\xi_{\mathbf{i}m}\xi_{\mathbf{j}n}$, then from Eq. (\ref{eq:representation-on-eigenvector-1-2}) it's easily obtained that the real-space correspondence of $\hat{P}_{g}\tilde{\Delta}^{s/t}_{mn}$ is $\Delta^{s/t}_{\hat{g}\mathbf{i},\hat{g}\mathbf{j}}$.  Therefore, if $\Delta_{\mathbf{ij}}$ is a real-space solution of the linearized gap equation with a $T_c$, then $\Delta_{\hat{g}\mathbf{i},\hat{g}\mathbf{j}}$ would also be a solution with the same $T_c$. As a result, all the real-space solutions $\{\Delta^{(\alpha)}_{\mathbf{ij}}\}$ corresponding to the same $T_c$ furnish an IR of the point group, expressed as,
\begin{equation}
\Delta_{\hat{g}\mathbf{i},\hat{g}\mathbf{j}}^{(\alpha)}=\sum_{\alpha'}D_{\alpha,\alpha'}^{(g)}\Delta_{\mathbf{i},\mathbf{j}}^{(\alpha')}.
\end{equation}
Here $D^{(g)}$ is the representation matrix corresponding to the element $g\in D_5$. From the IR that $\{D^{(g)}\}$ belongs to, we can classify the pairing symmetry of the group of pairing states with gap functions $\{\Delta^{(\alpha)}_{\mathbf{ij}}\}$.

In the remaining part of this section, we show the difference between the classifications of pairing symmetries on the QCs and periodic lattices: In the absence of the spin-orbit coupling (SOC) here, for each of the pairing symmetry on the QC, both spin-singlet and spin-triplet pairings are possible; while on centrosymmetric periodic lattices, the even (odd) orbital angular momentum is usually bound to the spin-singlet (triplet) pairing.

Usually, the classification of pairing symmetries is performed on the basis of the linearized gap equation or more general the linearized Ginzburg-Landau theory obtained just below $T_c$. At such temperatures, the pairing gap goes to zero, and the pairing state is in the weak-pairing limit (for most realistic superconductors, even the ground states belong to this limit). On periodic lattices in the weak-pairing limit, the Cooper pairing only takes place around the Fermi surface in the momentum space, and the Anderson's theorem requires that the pairing should only take place within intra-band. The pairing Hamiltonians are thus $H^{s}_{p} = \sum_{\mathbf{k}\alpha}(c_{\mathbf{k}\alpha\uparrow}^{\dagger}c_{-\mathbf{k}\alpha\downarrow}^{\dagger}-
    c_{\mathbf{k}\alpha\downarrow}^{\dagger}c_{-\mathbf{k}\alpha\uparrow}^{\dagger})\Delta_{\mathbf{k}}^{s,\alpha}+h.c.$ for the singlet pairings and $H^{t}_{p} =\sum_{\mathbf{k}\alpha}(c_{\mathbf{k}\alpha\uparrow}^{\dagger}c_{-\mathbf{k}\alpha\downarrow}^{\dagger}+
    c_{\mathbf{k}\alpha\downarrow}^{\dagger}c_{-\mathbf{k}\alpha\uparrow}^{\dagger})\Delta_{\mathbf{k}}^{t,\alpha}+h.c.$, or $\sum_{\mathbf{k}\alpha}c_{\mathbf{k}\alpha\uparrow}^{\dagger}c_{-\mathbf{k}\alpha\uparrow}^{\dagger}\Delta_{\mathbf{k}}^{t,\alpha}+h.c.$, or $\sum_{\mathbf{k}\alpha}c_{\mathbf{k}\alpha\downarrow}^{\dagger}c_{-\mathbf{k}\alpha\downarrow}^{\dagger}\Delta_{\mathbf{k}}^{t,\alpha}+h.c.$, or their arbitrary mixing
for the triplet ones, respectively. Here $\mathbf{k}$ and $\alpha$ label the momentum and band index respectively. Note that due to the Fermi statistics, the gap function $\Delta_{\mathbf{k}}^{s,\alpha}$ ($\Delta_{\mathbf{k}}^{t,\alpha}$) is even (odd) as function of $\mathbf{k}$, as its odd (even) part always cancels itself in the summation between $\mathbf{k}$ and $-\mathbf{k}$ in the $\sum_{\mathbf{k}}$. Generally, as an irreducible representation of the point group containing the inversion symmetry, the even (odd) gap function $\Delta_{\mathbf{k}}^{s,\alpha}$ ($\Delta_{\mathbf{k}}^{t,\alpha}$) belongs to the irreducible representation marked by even (odd) orbital angular momentum $l$. However, for a QC, the lattice momentum is no longer a good quantum number and the Anderson's theorem is broken, so there is no corresponding relationship between spin and orbital angular momenta.

Note that on noncentrosymmetric periodic lattice lack of inversion symmetry, each irreducible representation doesn't have definite parity of the pairing angular momentum. For example, on a lattice with $D_3$ symmetry, the pairing angular momentum $l=0$ or $l=1$ cannot be distinguished from $l=3-0=3$ or $l=3-1=2$, leading to ambiguity of the parity of the $l$. In such cases, the pairing angular momentum is also independent from the spin statistics.

\section{Topological Invariant}

Based on the $K$-theory, the Chern number for a finite-size system can be expressed as the following pseudo-spectrum
invariant index\cite{Loring2015, Fulga2016},
\begin{equation}
C_{ps}=\frac{1}{2}Sig\begin{pmatrix}X & Y+iH\\
Y-iH & -X
\end{pmatrix},\label{eq:c_ps}
\end{equation}
where $Sig$ represents the difference between the numbers of positive
and negative eigenvalues of the matrix acted on, $X$ and $Y$ are
position-operator matrices with $X$ defined by $X=\mathrm{diag}(x_{1},x_{1},x_{1},x_{1},\cdots,x_{N},x_{N},x_{N},x_{N})$ and $Y$ defined similarly. The BdG-Hamiltonian matrix $H$ is defined as
\begin{eqnarray}
H & = & \begin{pmatrix}h_{11} & h_{12} & \cdots & h_{1N}\\
h_{21} & h_{22} & \cdots & h_{2N}\\
\vdots & \vdots & \ddots & \vdots\\
h_{N1} & h_{n2} & \cdots\  & h_{NN}
\end{pmatrix},\label{eq:transformed-BdG}
\end{eqnarray}
in which $h$ is the matrix of local BdG-Hamiltonian defined as
\begin{eqnarray}
\mathcal{H}_{\mathrm{BdG}} & = & \sum_{ij}(c_{i\uparrow}^{\dagger}c_{i\downarrow}^{\dagger}c_{i\uparrow}c_{i\downarrow})h_{ij}(c_{j\uparrow}^{\dagger}c_{j\downarrow}^{\dagger}c_{j\uparrow}c_{j\downarrow})^{\dagger}.\label{eq:BdG}
\end{eqnarray}
From the above introduced definition of $C_{ps}$, it's shown below that $C_{ps}$ has two universal properties.

Firstly, any global unitary transformation will not change $C_{ps}$. Such unitary transformation is embodied as a site-independent unitary matrix $u$ performed on all the $h_{ij}$, leading to  $h_{ij}\to\tilde{h}_{ij}\equiv uh_{ij}u^{\dagger}$. Defining the unitary matrix $U=\mathrm{diag}(u,u,\cdots,u)$, it can be easily verified that
\begin{eqnarray}
U\begin{pmatrix}X & Y+iH\\
Y-iH & -X
\end{pmatrix}U^{\dagger} & = & \begin{pmatrix}X & Y+i\tilde{H}\\
Y-i\tilde{H} & -X
\end{pmatrix},
\end{eqnarray}
which suggests that the global unitary transformation will not change $C_{ps}$.

From the above property, it's easily known that the two TRS-breaking triplet pairing states have the same Chern number: one is described by the pairing term $\sum_{\mathbf{ij}}\Delta_{\mathbf{ij}}(c_{\mathbf{i}\uparrow}c_{\mathbf{j}\downarrow}+c_{\mathbf{i}\downarrow}c_{\mathbf{j}\uparrow})+h.c.$, i.e. the $(1,0)$ component; the other is described by $\sum_{\mathbf{ij}}\Delta_{\mathbf{ij}}(c_{\mathbf{i}\uparrow}c_{\mathbf{j}\uparrow}\pm c_{\mathbf{i}\downarrow}c_{\mathbf{j}\downarrow})+h.c.$, i.e. the $(1,1)\pm(1,-1)$ component. Obviously, the two chiral pairing states are mutually related by a global spin-SU(2) rotation (i.e. $\hat{s}_z\to \hat{s}_x$, for the case of $(1,1)-(1,-1)$) followed by a spin-dependent U(1)-gauge rotation (i.e. $c_{\downarrow}\to ic_{\downarrow}$, for the case of $(1,1)+(1,-1)$), which makes them share the same Chern number. Further more, the matrix in the right side of Eq.(\ref{eq:c_ps}) for the $(1,1)\pm(1,-1)$ state is block-diagonalized into the spin-up block and the spin-down one, suggesting that the Chern number of the two chiral triplet pairing state is twice as much as the one for a spinless system with the same pairing form factor.

Secondly, the $C_{ps}$ for a pairing state and that for its time-reversal (TR) conjugate is different by a sign. This property is proved as follow. Under TR transformation, we have
\begin{eqnarray}
Th_{ij}T^{-1} & = & \left(\begin{array}{cc}
-i\sigma_{y} & 0\\
0 & -i\sigma_{y}
\end{array}\right)h_{ij}^{*}\left(\begin{array}{cc}
i\sigma_{y} & 0\\
0 & i\sigma_{y}
\end{array}\right)=\left(\begin{array}{cc}
\sigma_{y} & 0\\
0 & \sigma_{y}
\end{array}\right)h_{ij}^{*}\left(\begin{array}{cc}
\sigma_{y} & 0\\
0 & \sigma_{y}
\end{array}\right)\nonumber \\
 & = & -\left(\begin{array}{cc}
\sigma_{y} & 0\\
0 & \sigma_{y}
\end{array}\right)\left(\begin{array}{cc}
0 & I_{2}\\
I_{2} & 0
\end{array}\right)h_{ij}\left(\begin{array}{cc}
0 & I_{2}\\
I_{2} & 0
\end{array}\right)\left(\begin{array}{cc}
\sigma_{y} & 0\\
0 & \sigma_{y}
\end{array}\right)=-\left(\begin{array}{cc}
0 & \sigma_{y}\\
\sigma_{y} & 0
\end{array}\right)h_{ij}\left(\begin{array}{cc}
0 & \sigma_{y}\\
\sigma_{y} & 0
\end{array}\right).
\end{eqnarray}
Here we have used the particle-hole symmetry of the BdG-Hamiltonian (in the third ``='' above). Since $\left(\begin{array}{cc} 0 & \sigma_{y}\\\sigma_{y} & 0\end{array}\right)$ is a site-independent unitary transformation, the $C_{ps}(H)$ is changed to $C_{ps}(-H)$ after the TR transformation. According to the following deriviation,
\begin{eqnarray}
Sig\begin{pmatrix}X & Y-iH\\
Y+iH & -X
\end{pmatrix} & = & -Sig\begin{pmatrix}-X & -Y+iH\\
-Y-iH & X
\end{pmatrix}=-Sig\left(\begin{array}{cc}
0 & I_{4N}\\
I_{4N} & 0
\end{array}\right)\begin{pmatrix}-X & -Y+iH\\
-Y-iH & X
\end{pmatrix}\left(\begin{array}{cc}
0 & I_{4N}\\
I_{4N} & 0
\end{array}\right)\nonumber \\
 & = & -Sig\begin{pmatrix}X & -Y-iH\\
-Y+iH & -X
\end{pmatrix}\nonumber \\
 & = & -Sig\left(\begin{array}{cc}
I_{4N} & 0\\
0 & -I_{4N}
\end{array}\right)\begin{pmatrix}X & -Y-iH\\
-Y+iH & -X
\end{pmatrix}\left(\begin{array}{cc}
I_{4N} & 0\\
0 & -I_{4N}
\end{array}\right)=-C_{ps}(H),
\end{eqnarray}
we know that $C_{ps}(H)=-C_{ps}(THT^{-1})$.

From the above property, we can know that the Chern numbers of all TRI pairing states are zero, because they should be equal to their opposite numbers. Therefore, all 1D-IR pairing states have zero Chern numbers, as the gap functions in these pairing states are real, which are TRI. Further more, the TRI helical triplet pairing state with pairing term $\sum_{\mathbf{ij}}(\Delta_{\mathbf{ij}}c_{\mathbf{i}\uparrow}c_{\mathbf{j}\uparrow}+ \Delta^{*}_{\mathbf{ij}}c_{\mathbf{i}\downarrow}c_{\mathbf{j}\downarrow})+h.c.$ and its gauge-rotated state $\sum_{\mathbf{ij}}(\Delta_{\mathbf{ij}}c_{\mathbf{i}\uparrow}c_{\mathbf{j}\uparrow}-\Delta^{*}_{\mathbf{ij}}c_{\mathbf{i}\downarrow}c_{\mathbf{j}\downarrow})+h.c.$ (with $c_{\downarrow}\to ic_{\downarrow}$) should also have zero Chern numbers.

Besides the above two universal properties, there is another important question to ask: for a triplet and a singlet pairing states with the same gap form factor except for the different exchanging parities, what's the relation between their Chern numbers? Obviously, the two states share the same pairing symmetry, but possess different total spins for the Cooper pair. Our numerical calculations on a lattice with 3466 sites and with restricted hopping integrals up to the third nearest neighbor show that such two pairing states share the same Chern numbers, although their total spins of Cooper pair are different, suggesting that the Chern number is only related to the pairing symmetry.

\section{The current operator}

The $\alpha$- component ($\alpha=x,y$) of the vectorial current operator $\hat{\mathbf{J}}_{\mathbf{i}}$  at a specific site $\mathbf{i}$ is defined as
\begin{equation}
\hat{J}_{\mathbf{i}\alpha}[\mathbf{A}]  =  -\frac{\delta\mathcal{\hat{H}}[\mathbf{A}]}{\delta A_{\mathbf{i}\alpha}} =  -\frac{\delta\mathcal{\hat{H}}_{\mathrm{TB}}[\mathbf{A}]}{\delta A_{\mathbf{i}\alpha}},\label{eq:current-operator-definition}
\end{equation}
where $\mathbf{A}$ is the vector potential, which appears in $\mathcal{\hat{H}}[\mathbf{A}]$ through a modification of the $\mathcal{\hat{H}}_{\mathrm{TB}}$ into
\begin{eqnarray}\label{H_TB_A}
\mathcal{\hat{H}}_{\mathrm{TB}}[\mathbf{A}] & = & \ensuremath{-\sum_{\mathbf{ij}\sigma}t_{\mathbf{ij}}\exp({i\int_{\mathbf{i}}^{\mathbf{j}}}\mathbf{A}\cdot d\mathbf{l})c_{\mathbf{i}\sigma}^{\dagger}c_{\mathbf{j}\sigma}}.
\end{eqnarray}
In our linear-response consideration of the superfluid density, the field $\mathbf{A}$ is a weak and smooth field. In such a limit, we can approximate Eq. (\ref{H_TB_A}) as
\begin{eqnarray}\label{H_TB_A_approx}
\mathcal{\hat{H}}_{\mathrm{TB}}[\mathbf{A}] & \approx & -\sum_{\mathbf{ij}\sigma}t_{\mathbf{ij}}[1+i(\mathbf{A}_{\mathbf{i}}+\mathbf{A}_{\mathbf{j}})\cdot\bm{R}_{\mathbf{ij}}/2\ensuremath{-[(\mathbf{A}_{\mathbf{i}}+\mathbf{A}_{\mathbf{j}})\cdot\mathbf{R}_{\mathbf{ij}})]^{2}/8]}c_{\mathbf{i}\sigma}^{\dagger}c_{\mathbf{j}\sigma}.
\end{eqnarray}
Substituting Eq. (\ref{H_TB_A_approx}) to Eq. (\ref{eq:current-operator-definition}), we obtain the formula of the $\alpha$- component of the current operator,
\begin{eqnarray}\label{current_operator}
\hat{J}_{\mathbf{i}\alpha}[\mathbf{A}] & = & \frac{i}{2}\sum_{\mathbf{l}\sigma}t_{\mathbf{il}}R_{\mathbf{il},\alpha}c_{\mathbf{i}\sigma}^{\dagger}c_{\mathbf{l}\sigma}+h.c.-\frac{1}{2}\sum_{\mathbf{l}\sigma}t_{\mathbf{il}}R_{\mathbf{il},\alpha}(\mathbf{R_{il}}\cdot\mathbf{A_i})c_{\mathbf{i}\sigma}^{\dagger}c_{\mathbf{l}\sigma}+h.c.,
\end{eqnarray}
where $R_{\mathbf{il},\alpha}$ is the $\alpha$- component of the relative position $\mathbf{R}_{\mathbf{il}}\equiv \mathbf{r}_{\mathbf{l}}-\mathbf{r}_{\mathbf{i}}$. The current operator consists of two parts, one is the constant part called as the paramagnetic current and the other is the part proportional to the vector potential called as the diamagnetic current.

Note that the current operator expressed as Eq.(\ref{current_operator}) is TR odd,
\begin{eqnarray*}
T\hat{\mathcal{\mathbf{J}}}_{\mathbf{i}}T^{-1} & = & -\hat{\mathbf{J}}_{\mathbf{i}}.
\end{eqnarray*}
Therefore, the expectation value $\langle\hat{\mathbf{J}}_{\mathbf{i}}\rangle$ in a TRI pairing state with $\mathbf{A}=0$ should always be zero. However, in a TRS-breaking chiral pairing state, it's possible to have  $\langle\hat{\mathbf{J}}_{\mathbf{i}}[\mathbf{A}=0]\rangle\ne 0$, suggesting the possibility of the spontaneous symmetry broken. But the total super current $\langle\sum_{\mathbf{i}}\hat{\mathbf{J}}_{\mathbf{i}}[\mathbf{A}=0]\rangle$ should be zero with $\mathbf{A}=0$, otherwise the macro superfluid density should be infinity.

Then, let's study the superfluid density, which is a tensor defined as
\begin{equation}
\rho_{s}^{\alpha\beta}\equiv\lim_{A_{\beta}\to0}\frac{-\langle A_{\beta}|\hat{J}_{\alpha}[A_{\beta}]|A_{\beta}\rangle}{A_{\beta}}.
\end{equation}
Here $\hat{J}_{\alpha}$ is the $\alpha$-component of the total super current per site defined as $\hat{\mathbf{J}}=\frac{1}{N}\sum_{\mathbf{i}}\hat{\mathbf{J}}_{\mathbf{i}}$. In the following, we shall prove that this tensor is diagonal and isotropic, i.e. $\rho^{\alpha\beta}_s=\rho_0\delta_{\alpha\beta}$, for all the pairing symmetries listed in the Table. I of the main text.

Let's consider an infinitesimal $\mathbf{A}$ imposed on the system, which will induce a super current $\mathbf{J}\equiv\langle\hat{\mathbf{J}}\rangle$, whose two components satisfy,
\begin{equation}
\left(\begin{array}{c}
J_{x}\\
J_{y}
\end{array}\right)=-\left(\begin{array}{cc}
\rho_{xx} & \rho_{xy}\\
\rho_{yx} & \rho_{yy}
\end{array}\right)\left(\begin{array}{c}
A_{x}\\
A_{y}
\end{array}\right).\label{eq:linear-response}
\end{equation}
Let's consider an arbitrary symmetry operation $g\in D_5$ acted on the system. For any pairing symmetry listed in the Table. I of the main text, the transformed gap function by $g$ can be expressed as $\hat{P}_{g}\Delta_{mn}\hat{P}_{g}^{-1}=e^{i\theta_{g}}\Delta_{mn}$, where $\theta_{g}$ is the angle according to the specific representation. Therefore, the transformed Hamiltonian $\hat{P}_{g}\hat{\mathcal{H}}_{\mathrm{BCS}}\hat{P}_{g}^{-1}$ can be recovered to the original one by a succeeding global gauge transformation, $c_{\mathbf{i},\sigma}\rightarrow e^{-i\theta_{g}/2}c_{\mathbf{i},\sigma}$, which will not change the superfluid density. As a result, after the $g\in D_5$ symmetry operation acted on the system, the superfluid density tensor $\left(\begin{array}{cc} \rho_{xx} & \rho_{xy}\\ \rho_{yx} & \rho_{yy} \end{array}\right)$ doesn't alert. On the other hand, the vectorial current and vector potential are transformed as
\begin{equation}
\left(\begin{array}{c}
J_{x}\\
J_{y}
\end{array}\right)\rightarrow R_{g}\left(\begin{array}{c}
J_{x}\\
J_{y}
\end{array}\right),\left(\begin{array}{c}
A_{x}\\
A_{y}
\end{array}\right)\rightarrow R_{g}\left(\begin{array}{c}
A_{x}\\
A_{y}
\end{array}\right).
\end{equation}
As a result, we have $[\rho_s,R_{g}]=0$ for any $g\in D_5$. There are two types of $R_{g}$,
\begin{equation}
R_{\hat{C}_{\frac{2\pi n}{5}}}=\left(\begin{array}{cc}
\cos\theta_{n} & -\sin\theta_{n}\\
\sin\theta_{n} & \cos\theta_{n}
\end{array}\right)=\cos \theta_{n} I-i\sin \theta_{n}\sigma_y,R_{\hat{\sigma}_{x}}=\left(\begin{array}{cc}
1 & 0\\
0 & -1
\end{array}\right)=\sigma_z.
\end{equation}
Setting $\rho_s=\alpha I+\beta\sigma_{x}+\mu\sigma_{y}+\nu\sigma_{z}$, it can be verified that $[\rho_s,R_{g}]=0 \to \beta=\mu=\nu=0$. As a result, we have $\rho^{\alpha\beta}_s=\rho_0\delta_{\alpha\beta}$. Our numerical calculation agrees with this theoretical result, and the linear response along the direction $x$ is shown in Fig. \ref{fig:linear_response}, which suggests a nonzero superfluid density.
\begin{figure}[h]
\includegraphics[width=0.48\textwidth]{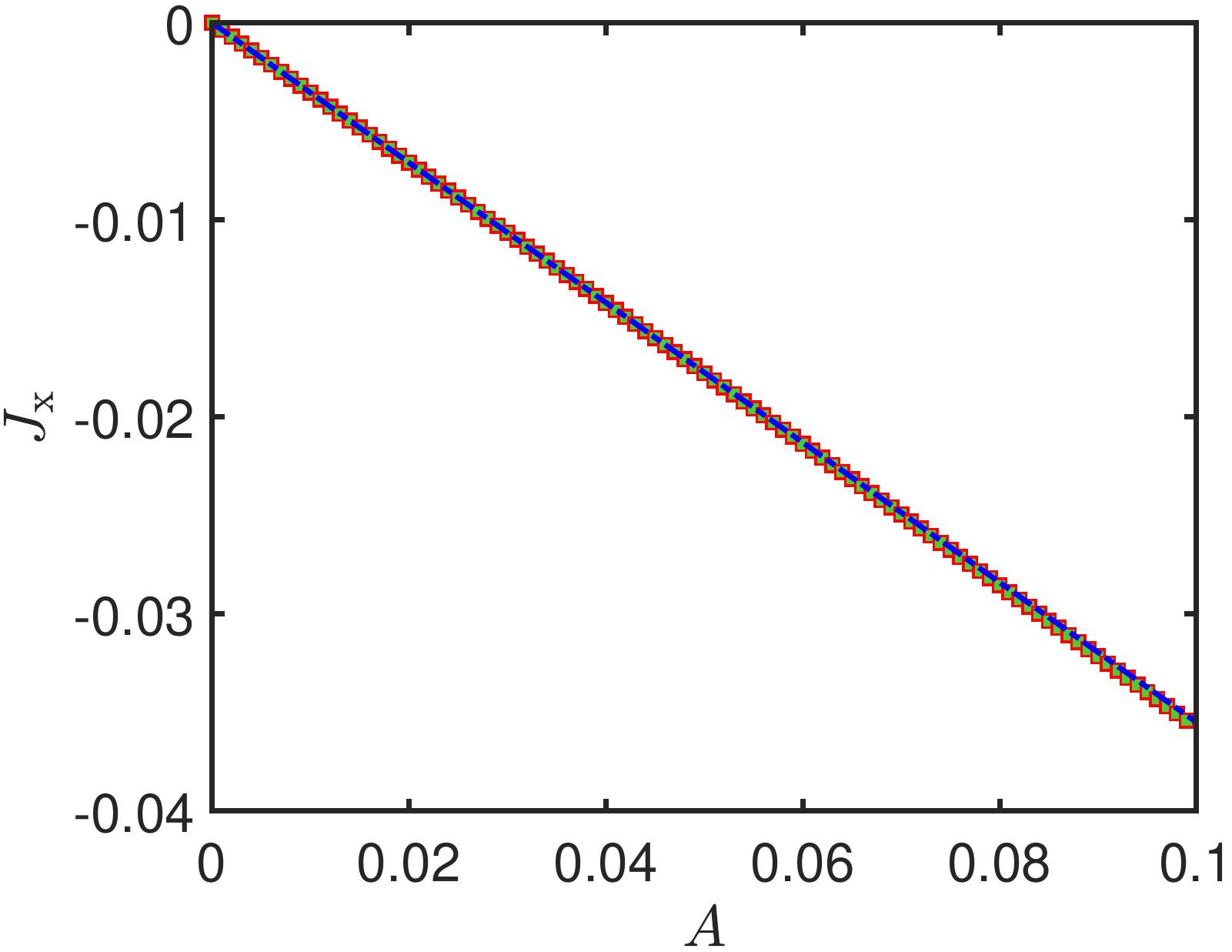}
\caption{\label{fig:linear_response}(Color online) Linear response of the
macro super current to an imposed vector potential $\mathbf{A}$ along the $x$- direction
in a triplet $d+id/f+if$ state at the  filling 0.985 and $U/W_{D}\text{\ensuremath{\approx}}0.09$.}

\end{figure}

\end{widetext}


\begin{thebibliography}{10}
\bibitem{Goldman} See, A. I. Goldman and R. F. Kelton, Rev. Mod. Phys. \textbf{65}, 213 (1993) and the references there.

\bibitem{Shechtman} D. Shechtman, I. Blech, D. Gratias, and J. W. Cahn, Phys. Rev. Lett. \textbf{53}, 1951 (1984).

\bibitem{Tsunetsugu1} H. Tsunetsugu, T. Fujiwara, K. Ueda, T. Tokihiro, Phys. Rev. B \textbf{43}, 8879 (1991).

\bibitem{Tsunetsugu2} H. Tsunetsugu, K. Ueda, Phys. Rev. B \textbf{43}, 8892 (1991).

\bibitem{Susumu} S. Yamamoto and T. Fujiwara, Phys. Rev. B \textbf{51}, 8841 (1995).

\bibitem{Wessel} S. Wessel, A. Jagannathan, and S. Haas, Phys. Rev. Lett. \textbf{90}, 177205 (2003).

\bibitem{Thiem} S. Thiem and J. T. Chalker, Phys. Rev. B \textbf{92}, 224409 (2015).

\bibitem{Koga} A. Koga and H. Tsunetsugu, Phys. Rev. B \textbf{96}, 214402 (2017).

\bibitem{Otsuki} J. Otsuki and H. Kusunose, J. Phys. Soc. Jap. \textbf{85}, 073712 (2016).

\bibitem{Watanabe} S. Watanabe and K. Miyake, J. Phys. Soc. Jap. \textbf{85}, 063703 (2016).

\bibitem{Shaginyan} V. R. Shaginyan, A. Z. Msezane, K. G. Popov, G. S. Japaridze, and V. A. Khodel, Phys. Rev. B \textbf{87}, 245122 (2013).

\bibitem{Takemori} N. Takemori and A. Koga, J. Phys. Soc. Jap. \textbf{84}, 023701 (2015).

\bibitem{Takemura} S. Takemura, N. Takemori, and A. Koga, Phys. Rev. B \textbf{91}, 165114 (2015).

\bibitem{Andrade} E. C. Andrade, A. Jagannathan, E. Miranda, M. Vojta, and V. Dobrosavljevic, Phys. Rev. Lett. \textbf{115}, 036403 (2015).

\bibitem{Kraus} Y. E. Kraus, Y. Lahini, Z. Ringel, M. Verbin, and O. Zilberberg, Phys. Rev. Lett. \textbf{109}, 106402 (2012).

\bibitem{Huang1} H. Huang and F. Liu, Phys. Rev. Lett. \textbf{121}, 126401 (2018).

\bibitem{Huang2}H. Huang and F. Liu, Phys. Rev. B \textbf{100}, 085119 (2019).

\bibitem{Longhi} S. Longhi, Phys. Rev. Lett. \textbf{122}, 237601 (2019).

\bibitem{Autti} S. Autti, V. B. Eltsov, and G. E. Volovik, Phys. Rev. Lett. \textbf{120}, 215301 (2018).

\bibitem{Giergiel} K. Giergiel, A. Kuro\ifmmode \acute{s}\else\'{s}\fi{}, and K. Sacha, Phys. Rev. B \textbf{99}, 220303 (2019).

\bibitem{Lang} L.-J. Lang, X. Cai, and S. Chen, Phys. Rev. Lett. \textbf{108}, 220401 (2012).

\bibitem{Sanchez} L. Sanchez-Palencia and L. Santos, Phys. Rev. A \textbf{72}, 053607 (2005).

\bibitem{Singh} K. Singh, K. Saha, S. A. Parameswaran, and D. M. Weld, Phys. Rev. A \textbf{92}, 063426 (2015).

\bibitem{Bandres2016} M. A. Bandres, M. C. Rechtsman, and M. Segev, Phys. Rev. X \textbf{6},  011016 (2016).

\bibitem{Hou} J. Hou, H. Hu, K. Sun, and C. Zhang, Phys. Rev. Lett. \textbf{120}, 060407 (2018).

\bibitem{Varjas2019} D. Varjas, A. Lau, K. P\"oyh\"onen, A. R. Akhmerov, D. I. Pikulin, and I. C. Fulga, Phys. Rev. Lett. \textbf{123}, 196401 (2019).

\bibitem{Spurrier2020} S. Spurrier and N. R. Cooper, arXiv:2001.05511.

\bibitem{exp} K. Kamiya, T. Takeuchi, N. Kabeya, N. Wada, T. Ishimasa, A. Ochiai, K. Deguchi, K. Imura, and N. K. Sato, Nat. Commun. \textbf{9}, 154 (2018).

\bibitem{exp2} K. M. Wong, E. Lopdrup, J. L. Wagner, Y. Shen, and S. J. Poon, Phys. Rev. B \textbf{35}, 2494 (1987).

\bibitem{exp3} J. L. Wagner, B. D. Biggs, K. M. Wong, and S. J. Poon, Phys. Rev. B \textbf{38}, 7436 (1988).

\bibitem{exp4} K. Deguchi, M. Nakayama, S. Matsukawa, K. Imura, K. Tanaka, T. Ishimasa, and N. K. Sato, J. Phys. Soc. Jap. \textbf{84}, 023705 (2015).

\bibitem{Sakai2017} S. Sakai, N. Takemori, A. Koga, and R. Arita, Phys. Rev. B \textbf{95}, 024509 (2017).


\bibitem{theory1} R. N. Ara\'ujo and E. C. Andrade, Phys. Rev. B \textbf{100}, 014510 (2019).

\bibitem{theory2} S. Sakai and R. Arita, Phys. Rev. Res. \textbf{1}, 022002(R) (2019).

\bibitem{theory3} Y. Nagai, arXiv:2001.02362.

\bibitem{attractive} Y.-Y. Zhang, Y.-B. Liu, Y. Cao, W.-Q. Chen and F. Yang, preprint

\bibitem{Cooper_instability} L. N. Cooper, Phys. Rev. \textbf{104}, 1189 (1956).

\bibitem{Anderson} P. W. Anderson, J. Phys. Chem. Solids. \textbf{11}, 26 (1959).

\bibitem{KL1} W. Kohn and J. M. Luttinger, Phys. Rev. Lett. \textbf{15}, 524 (1965).

\bibitem{KL2} M. A. Baranov, A. V. Chubukov, and M. Yu. Kagan, Int. J. Mod. Phys. B \textbf{06}, 2471 (1992).

\bibitem{Penrose1974} R. Penrose, Bull. Inst. Math. Appl. \textbf{10}, 266 (1974).

\bibitem{footnote1} Note that here we have set the center of the Perose lattice as the coordinate origin, and taken a symmetry-respected boundary to reflect the thermal-dynamic limit behavior.

\bibitem{footnote2} In the calculation of the DOS, we count the number $N_E$ of the states locating within a narrow energy shell $\Delta_E$
near $E_F$, and the DOS is obtained as $\frac{N_E}{N\Delta_E}$, where $N$ is the site number. Tune $N$ and $\Delta_E$, until a converged result of the DOS is obtained.

\bibitem{Supplementary} See the Appendix for our real-space perturbative theory on the repulsive Hubbard model, the basis for the classification of pairing symmetries on the QC, the definition and properties of the topological invariant and the current operator.

\bibitem{footnote3} On a centrosymmetric periodic lattice without SOC, the pairing angular momentum $l$ and the spin statistics are mostly mutually determined\cite{Sigrist}, particularly in the intra-band pairing case\cite{Qixiaoliang}. On a noncentrosymmetric lattice, a given irreducible representation of the point group doesn't possess definite parity of $l$. In this case, the parity of $l$ is also independent from the spin statistics.

\bibitem{Sigrist} M. Sigrist and K. Ueda, Rev. Mod. Phys. \textbf{63}, 239 (1991).

\bibitem{Qixiaoliang} X.-L. Qi, T.- L. Hughes and S.-C. Zhang, Phys. Rev. B \textbf{81},134508 (2010).

\bibitem{Loring2015}  T. A. Loring, Ann. Phys. (N. Y). \textbf{356}, 383 (2015).

\bibitem{Fulga2016} I. C. Fulga,  D. I. Pikulin, and T. A. Loring, Phys. Rev. Lett. \textbf{116}, 257002 (2016).

\bibitem{Horovitz2003} B. Horovitz and A. Golub, Phys. Rev. B \textbf{68}, 214503 (2003).

\bibitem{Stone2004}  M. Stone and R. Roy, Phys. Rev. B \textbf{69}, 184511 (2004).

\bibitem{Wang2005} B. Braunecker, P. A. Lee, and Z. Wang, Phys. Rev. Lett. \textbf{95}, 017004 (2005).

\bibitem{Sauls2011} J. A. Sauls, Phys. Rev. B \textbf{84}, 214509 (2011).

\bibitem{Kallin2012} C. Kallin, Rep. Prog. Phys. \textbf{75}, 042501 (2012).

\bibitem{Huang2014} W. Huang, E. Taylor, and C. Kallin, Phys. Rev. B \textbf{90}, 224519 (2014).

\bibitem{Liu2004} K. D. Nelson, Z. Q. Mao, Y. Maeno, and Y. Liu, Science \textbf{306}, 1151 (2004).

\bibitem{Kirtley2007} J. R. Kirtley, C. Kallin, C. W. Hicks, E.-A. Kim, Y. Liu, K. A. Moler, Y. Maeno, and K. D. Nelson, Phys. Rev. B \textbf{76}, 014526 (2007).

\bibitem{Curran2014} P. J. Curran, S. J. Bending, W. M. Desoky, A. S. Gibbs, S. L. Lee, and A. P. Mackenzie, Phys. Rev. B \textbf{89}, 144504 (2014).

\bibitem{TBG1} S. J. Ahn, P. Moon, T.-H. Kim, H.-W. Kim, H.-C. Shin, E. H. Kim, H. W. Cha, S.-J. Kahng, P. Kim, M. Koshino, Y.-W. Son, C.-W. Yang, and J. R. Ahn, Science, \textbf{361}, 782 (2018).

\bibitem{TBG2}W. Yao, E. Wang, C. Bao, Y. Zhang, K. Zhang, K. Bao, C. K. Chan, C. Chen, J. Avila, M. C. Asensio, J. Zhu, and S. Zhou, Proc. Natl. Acad. Sci. \textbf{115}, 6928 (2018).

\end{thebibliography}

\begin{thebibliography}{1}
\bibitem{KL1} W. Kohn and J. M. Luttinger, Phys. Rev. Lett. \textbf{15}, 524 (1965).
\bibitem{KL2} M. A. Baranov, A. V. Chubukov, and M. Yu. Kagan, Int. J. Mod. Phys. B \textbf{06}, 2471 (1992).
\bibitem{Mahan} Gerald D. Mahan, ``Many-Particle Physics (Second Edition)'', Plenum Press, New York and London (1990).
\bibitem{Loring2015} T. A. Loring, Ann. Phys. (N. Y). \textbf{356},
383 (2015).
\bibitem{Fulga2016} I. C. Fulga,  D. I. Pikulin, and T. A. Loring, Phys. Rev. Lett. 116, 257002 (2016).
\end{thebibliography}
\end{document}